\documentclass[fleqn,usenatbib]{mnras2}

\usepackage{newtxtext,newtxmath}

\usepackage{pdflscape}

\usepackage[T1]{fontenc}

\DeclareRobustCommand{\VAN}[3]{#2}
\let\VANthebibliography\thebibliography
\def\thebibliography{\DeclareRobustCommand{\VAN}[3]{##3}\VANthebibliography}

\usepackage{graphicx}	
\usepackage{amsmath}	
\usepackage{newtxtext,newtxmath}
\usepackage{tikz} 
\usepackage{verbatim}
\usepackage{booktabs}       
\usepackage{here}

\usepackage{hyperref}
\hypersetup{
    colorlinks=true,
    linkcolor=blue,
    filecolor=magenta,      
    urlcolor=cyan,
}

\graphicspath{{Figures/}}
\newlength{\fullwidth}
\newlength{\halfwidth}
\def\checkmark{\tikz\fill[scale=0.4](0,.35) -- (.25,0) -- (1,.7) -- (.25,.15) -- cycle;} 

\title[\textsc{CRESCENDO}]{\textsc{CRESCENDO}: An on-the-fly Fokker-Planck Solver for Spectral Cosmic Rays in Cosmological Simulations}

\author[L. M. Böss et al]{
Ludwig M. Böss$^{1,2}$\thanks{E-mail: lboess@usm.lmu.de},
Ulrich P. Steinwandel$^{3}$,
Klaus Dolag$^{1,4}$,
Harald Lesch$^{1}$
\\
$^{1}$Universit\"ats-Sternwarte München, Fakultät für Physik, Ludwig-Maximilians Universit\"at M\"unchen, Scheinerstr. 1, 81679 Munich, Germany\\
$^{2}$Excellence Cluster ORIGINS, Boltzmannstr. 2, 85748, Garching, Germany\\
$^{3}$Center for Computational Astrophysics, Flatiron Institute, 162 5th Avenue, New York, NY 10010, USA\\
$^{4}$Max Planck Institute for Astrophysics, Karl-Schwarzschild-Str. 1, 85741 Garching, Germany
}

\date{Accepted 2022 December 1. Received 2022 November 17; in original form 2022 July 11}

\pubyear{2022}

\begin{document}
\label{firstpage}
\pagerange{\pageref{firstpage}--\pageref{lastpage}}
\maketitle

\begin{abstract}
Non-thermal emission from relativistic Cosmic Ray (CR) electrons gives insight into the strength and morphology of intra-cluster magnetic fields, as well as providing powerful tracers of structure formation shocks.
Emission caused by CR protons on the other hand still challenges current observations and is therefore testing models of proton acceleration at intra-cluster shocks.
Large-scale simulations including the effects of CRs have been difficult to achieve and have been mainly reduced to simulating an overall energy budget, or tracing CR populations in post-processing of simulation output and has often been done for either protons or electrons.
We introduce \textsc{CRESCENDO}: \textit{Cosmic Ray Evolution with SpeCtral Electrons aND prOtons}, an efficient on-the-fly Fokker-Planck solver to evolve distributions of CR protons and electrons within every resolution element of our simulation. The solver accounts for CR (re-)acceleration at intra-cluster shocks, based on results of recent PIC simulations, adiabatic changes and radiative losses of electrons. We show its performance in test cases as well as idealized galaxy cluster (GC) simulations.
We apply the model to an idealized GC merger following best-fit parameters for \textsc{CIZA J2242.4+5301-1} and study CR injection, radio relic morphology, spectral steepening and synchrotron emission.
\end{abstract}

\begin{keywords}
Physical data and processes: cosmic rays -- acceleration of particles -- MHD -- plasmas -- methods: numerical -- galaxies: clusters
\end{keywords}



\section{Introduction}

The Intracluster medium (ICM) can be characterised as a low density ($n \sim 10^{-3}$ cm$^{-3}$), high temperature ($T \sim 10^{8}$K, E > 1 keV), strongly ionised, weakly collisional, high-$\beta$ ($\beta \equiv P_\mathrm{th}/P_\mathrm{mag} \sim 100$, with $B \sim 0.1-10 \mu$G) plasma \citep[e.g.][]{Carilli2002}.
Radio observations of Galaxy Clusters indicate the presence of relativistic electrons (Cosmic Ray electrons, CRe) in the ICM.
These observations of the extended radio emission around galaxy clusters are paramount to obtain constraints on the cluster magnetic field strength \citep[e.g.][]{Clarke2001, Brueggen2012, Johnston2015}. 
Galaxy cluster magnetic fields on the other hand are of outstanding importance for particle acceleration mechanisms which are likely the origin of two of the most common flavours of radio emission in merging galaxy clusters, radio haloes (RH) and radio relics (RR).
Both types of radio sources are extensively studied in the literature \citep[see e.g.][for reviews]{Feretti2012, Brueggen2012, Weeren2019}.\\
RRs can be characterized as elongated structures observable in radio wavelengths, that are most likely formed due to cluster merger-shocks in the outskirts of the ICM.
They can show strong polarisation and sharp edges, indicating strong magnetic fields and efficient acceleration of electrons.
How the mechanisms of acceleration from a thermal pool of particles to a relativistic population works in detail for these systems has been a matter of some debate in the literature.
The efficient acceleration of particles at shocks can be described via Diffusive Shock Acceleration (DSA) \citep[e.g.][]{Bell_1978a, Bell_1978b, Blandford1987}.
Within DSA particles repeatedly cross a shock front and are reflected up- and downstream by scattering off electromagnetic turbulence \citep[see e.g.][for reviews]{Drury1983, Bykov2019}.
As this process is self-similar the particle distribution naturally approaches a power-law in energy or momentum.
However, the main complication with this process is that it is only effective once the gyroradius of the particle is comparable to the shock width and the particle can gain energy over multiple cycles.
Especially for electrons this requires the particles to gain substantial energy before they can be efficiently accelerated by DSA.
Otherwise they simply cross the shock, gain energy and are advected downstream before they can be scattered into the upstream again.
Small scale simulations with particle-in-cell (PIC) and hybrid models have been used to find solutions for these problems \citep[e.g.][]{Caprioli2014, Guo2014, Park2015, Caprioli2018, Ryu2019, Kobzar2021, Ha2021}.
It has been shown that particles can initially gain energy via (stochastic) shock drift acceleration (sSDA), where they gain energy from the gradient drift along the shock ramp until they reach a critical momentum $p_\mathrm{inj}$ which is enough to inject them into a Fermi-like process like DSA \citep[e.g.][and references therein]{Ha2021}.
The onset of sSDA (and with that DSA) is found to depend on the excitement of plasma instabilities and electromagnetic waves which require a critical sonic Mach number $\mathcal{M}_\mathrm{s,crit} \approx 2.25$ \citep[][]{Ha2018}, and different magnetic field configurations to be triggered \citep[see e.g.][for a summary of the effects]{Ha2021}.
In this context it was found that CR protons are more efficiently accelerated by low obliquity shocks \citep[e.g.][]{Caprioli2014, Caprioli2018, Ryu2019}, meaning shocks where the angle between magnetic field vector and shock normal are small, while electrons are more efficiently accelerated at high obliquity shocks \citep[e.g.][]{Guo2014, Ha2021, Kobzar2021}.
However, see \citet{Winner2020, Shalaby2021, Shalaby2022} for discussions of this in the case of low-$\beta$ Supernova (SN) remnants.
The acceleration efficiencies (the fraction of available energy dissipated by the shock that goes into the acceleration of CRs) required to reproduce the radio surface brightness of radio relics should also inject a significant population of CR protons.
These should then interact with background gas and scatter into $\pi^0$ pions and from there $\gamma$ photons, which should be observable \citep[see e.g.][for a recent review of this problem]{Wittor2021}.
This is ruled out by FERMI observations which place an upper limit to the CR proton energy density in clusters at a few per cent of the thermal energy density \citep[e.g.][and references therein]{Ackermann2014, Ackermann2015, Ackermann2016, Vazza2016, Adam2021}. 
One possible solution to study this problem in simlations that has been proposed is the inclusion of shock obliquity in the acceleration efficiency models \citep[][]{ha_gamma-ray_2020, Wittor_2020_limit_Protons}.\\
RHs on the other hand can be classified as structures in which the diffuse radio emission is following the thermal structure of the ICM (i.e. the X-ray emitting hot gas).
They have sizes of up to 2 Mpc that largely follow the overall structure imprinted on ICM-scales \citep[for example in the Coma cluster][]{Large1959, Willson1970, Giovannini1993, Thierbach2003, Brown2011, Bonafede2022} and can be detected out to large redshifts \citep[e.g, El-Gordo][]{Menanteau2012}.
The spectral index of the integrated synchrotron emission are in good agreement with the power-law index range of $\alpha \approx -1.4 \sim -1.1 $ \citep[e.g.][]{Giovannini2009}.
While the underlying physical origin is still under debate, there is strong observational evidence that these haloes correlate with recent merger activity \citep[][and references therein]{Cassano2010} which can lead to an increase in the turbulent motions in the ICM and with that re-acceleration of CRes \citep[e.g.][]{Cassano2005, Brunetti2007, Brunetti2016, Brunetti2016a, Eckert2017, Stuardi2019, Wong2020}.
Another possible explanation is a hadronic origin, where long-lived CR protons produce secondary electrons \citep[e.g.][]{Dolag2000, Pfrommer2004}, which could explain observations like the unbroken spectral index up to very high energies in the halo center observed by \citet{Perrott2021}.\\
Even state-of-the-art galaxy cluster simulations lack the resolution to model these processes from first principle.
However, CR-injection at supersonic shocks can be implemented by adopting mach number dependent efficiency models obtained in thermal leakage models, PIC or hybrid simulations \citep[e.g.][]{Kang2007_model, Kang2013, Caprioli2014, Ryu2019} and bridging the gap between PIC and large scale MHD simulations will be one of the challenges of the upcoming simulation sets including CR physics in galaxy clusters \citep[for simulations of intermediate scales see e.g.][]{Vaidya2018, Dominguez-Fernandez2021, Dominguez-Fernandez2021a}.
Large-scale simulations that study CRs can broadly be separated into two categories.
The first is pure post-processing of dynamically decoupled CRs.
These CRs can be modelled as a single energy budget following a strict power-law distribution in energy and have been used to study galaxy cluster radio haloes and intra-cluster shocks \citep[e.g.][]{Ensslin1998, Dolag2000, Kang2007, Hoeft2007, Hoeft2008, Hong2014, Hong_2015, Wittor2017, Banfi2020, ha_gamma-ray_2020}.
As processes such as energy losses of CRs are energy dependent, a simple power-law approach can be limiting the descriptive capability of these models.
To this end the model can be extended to describe a population of CR electrons and protons, and their distribution function can be evolved in time using a Fokker-Planck solver \citep[][]{Pinzke2013, Pinzke2016, Donnert2014a, Winner2019, Winner2020, Vazza2021}.
These post-processing approaches are however limited in information by the number of output snapshots or information of tracer particles in the simulation and interpolations between those outputs.\\
The second category is an on-the-fly implementation of CRs.
This requires a large surrounding code infrastructure with descriptions for shock finding, star formation, or AGNs as CR sources, as well as an accurate treatment of turbulence and magnetic fields.
For these reasons it has been a significant computational challenge.
To reduce the computational cost of the CR component itself it has mainly been treated as an additional energy budget coupled to the hydrodynamical equations as an ideal, relativistic gas.
This approach is also referred to as \textit{one bin approach} or \textit{gray model} \citep[e.g.][]{Girichidis2016} and has been used to model the impact of CRs on galaxy formation \citep[e.g.][]{Hanasz2003, Jubelgas2008, Girichidis2016, Pfrommer2016, Ruszkowski2017, Butsky2018, Butsky2020, Kim2020, Semenov2021, Chan2019, Chan2021, Weber2022}, AGN jets \citep[e.g.][]{Sijacki2008, Guo2011, Ruszkowski2017a} and structure formation \citep[e.g.][]{enslin_cosmic_2007, Pfrommer2007, Pfrommer2008, Pfrommer2016, Vazza2012, Vazza2016}.\\
As in the post-processing models this one-bin approach can be extended to account for energy dependent processes by evolving a distribution function of CRs in time.
To this end \citet{Jones1999} extended the treatment of CR electrons in their work to represent a population of particles distributed in momentum space following a piece-wise power-law, while \citet{Miniati2001, Miniati2001a} extended this further to a spectral model for both protons and electrons.
This allows them to more accurately model fast radiative loss processes of electrons as well as adiabatic changes, injection and propagation, as they are not limited by the time resolution of the outputs.
\citet{Jones2005} further improved upon this, including CR propagation in a method they labelled "Coarse-Grained Momentum finite Volume" (CGMV).
More recently \citet{Yang2017, Yang2018, Girichidis2020, Girichidis2022, Ogrodnik2020, Hopkins2021a} revisited spectral CR models for novel implementations in current state of the art cosmological MHD codes used in simulations of galaxy formation.\\
The goal of this work is to introduce a novel implementation of an on-the-fly Fokker-Planck solver for both CRp and CRe to study galaxy clusters.
We will show its practical applicability to study RRs in simulations of idealized galaxy cluster mergers before we apply the model to MHD simulations of cosmological structure formation to study RRs and RHs in future work.
This paper is structured as follows: In Section \ref{sec:model} we introduce the spectral CR model and the relevant physical processes for this work.
Section \ref{sec:tests} shows a number of tests of the accuracy and performance of the model.
In Section \ref{sec:cluster_mergers} we apply the model to simulations of idealized galaxy cluster mergers and discuss the impact on the simulation, as well as the observables that can be obtained for the northern relic (Section \ref{sec:NR}) and the southern relic (Section \ref{sec:SR}).
Section \ref{sec:conclusion} sums up our results and gives an outlook to future work.
\section{Cosmic Ray Model}
\label{sec:model}
Our aim is to study the impact of CR protons on structure formation processes and obtain observables from CR electrons. As the typical resolution elements of cosmological simulations are 60-70 orders of magnitude above individual proton and electron masses, we need to implement a sub-grid model to treat whole populations of these particle species.
In principle CRs are distributed in phase-space according to their momentum $\mathbf{p}$ and their position $\mathbf{x}$ at time $t$ in the distribution function $F(\mathbf{p}, \mathbf{x}, t)$. Assuming a sufficiently stochastic \citep{Drury1983} scattering process, e.g. the scattering and self-confinement by Alfvén waves triggered by the CR streaming instability \citep[e.g.][]{Kulsrud1969, Wentzel1974, Skilling1975_I, Skilling1975_II, Skilling1975_III}, we can infer that the particle movement is random on small scales and treat the distribution of the CRs as isotropic in momentum space. This simplifies the phase-space from $F(\mathbf{p}, \mathbf{x}, t)$, to only depend on absolute momenta
\begin{align}
	F(\mathbf{p}, \mathbf{x}, t) \rightarrow 4\pi p^2 f(p, \mathbf{x}, t)
	\label{eq:distr_func_simplification}
\end{align}
The time evolution of this distribution function can then be described by the diffusion-advection equation \citep[see e.g.][for a derivation]{Skilling1975_I, Drury1983, Schlickeiser2002}
\begin{align}
	\frac{D f(p,\mathbf{x},t)}{Dt} = \nabla \left( \kappa(p) \nabla f(p,\mathbf{x},t) \right) \label{eq:fp-xdiffusion}
\end{align}
\begin{align}
	\quad \quad \quad \quad + \left( \frac{1}{3} \nabla \cdot \mathbf{u} \right) p \frac{\partial f(p,\mathbf{x},t)}{\partial p} \label{eq:fp-adiabatic} 
\end{align}
\begin{align}
	\quad \quad \quad \quad + \frac{1}{p^2} \frac{\partial}{\partial p } \left( p^2 \left[ \sum_l b_l f(p,\mathbf{x},t) + D_{\mathrm{pp}} \frac{\partial f(p,\mathbf{x},t)}{\partial p} \right] \right) \label{eq:fp-rad}
\end{align}
\begin{align}
	\quad \quad \quad \quad - \frac{f(p,\mathbf{x},t)}{t_c(p)} \label{eq:fp-catastrophic}
\end{align}
\begin{align}
	\quad \quad \quad \quad + j(\mathbf{x}, p, t), \label{eq:fp-sources}
\end{align}
where the individual terms describe;
Advection (l.h.s. side of Eq.~\ref{eq:fp-xdiffusion}), which we express in Lagrangian form by using: $\frac{Df}{Dt} = \frac{\partial f}{\partial t} + \mathbf{u} \cdot \nabla f$,
spatial diffusion (r.h.s of Eq.~\ref{eq:fp-xdiffusion}), adiabatic compression/expansion of the surrounding gas (Eq.~\ref{eq:fp-adiabatic}), energy losses (first term of Eq.~\ref{eq:fp-rad}) where we used
\begin{equation}
    \sum_l b_l \equiv \left\vert \frac{dp}{dt} \right\vert_c + \left\vert \frac{dp}{dt}\right\vert_\mathrm{rad}
\end{equation}
for energy losses due to coulomb interaction and radiative losses, respectively. Diffusion in momentum space, or Fermi-II re-acceleration is described by the second term of Eq.~\ref{eq:fp-rad}.
Catastrophic losses are given via Eq.~\ref{eq:fp-catastrophic} and sources of CRs are described by Eq.~\ref{eq:fp-sources}.
For the remainder of this work we will focus on adiabatic changes, radiative losses and shocks as CR sources. We will introduce the treatments of further terms in future work where applicable.
\subsection{Spectral Parameters}
In order to evolve Eq.~\ref{eq:fp-xdiffusion} in time, we need to choose a numerical discretisation.
Similar to \citet{Yang2017, Girichidis2020, Ogrodnik2020} we follow the approach by \citet{Miniati2001} and parameterize the spectrum with four parameters: For a momentum bin $p_i$ we follow the spectral norm $f_i$, spectral slope $q_i$, CR number $N_i$ and CR energy $E_i$. Since observations \citep[e.g.][]{Abdo2009, Aguilar2013} show the CR spectrum as a (broken) power-law in energy or momentum over many orders of magnitude, a logical choice is to discretize the initial spectrum as a power-law in momentum space split up in $i$ momentum bins of equal size. This way the spectrum takes the functional form
\begin{align}
	f(p) = f_i \left(\frac{p}{p_i}\right)^{-q_i}
	\label{eq:piecewise_powerlaw}
\end{align}
where $f_i$ and $q_i$ are norm and slope of the spectrum at momentum $p_i$.
Compared to a piece-wise constant discretisation this piece-wise powerlaw approach has the additional advantage of spanning the high dynamical range with fewer momentum bins and thus saving on memory, while maintaining a better representation of the spectral shape.
Once this discretisation has been set we can obtain the number of CRs per bin by performing a simple volume-integral in momentum space in shells where the width corresponds to the bin-widths.
\begin{align}
	N_i = \frac{1}{\rho} \int\limits_{p_i}^{p_{i+1}} dp \: 4 \pi p^2 f(p)
		= \frac{4 \pi f_i p_i^3}{\rho } \frac{\left( \left( \frac{p_{i+1}}{p_i}\right)^{3-q_i} - 1 \right)}{3 - q_i } \label{eq:n_i_analytic}
\end{align}
The energy contained in each bin can be obtained by assuming that every CR particle carries the kinetic energy 
\begin{equation}
T(p) =  \sqrt{p^2c^2 + m_{i}^2 c^4} - m_{i} c^2 \approx pc    
\end{equation}
where we used that the CRs we are interested in are relativistic and $m_i$ represents the rest mass of the individual particles species. Including this in our volume integral gives
\begin{align}
	E_i = \frac{1}{\rho} \int\limits_{p_i}^{p_{i+1}} dp \: 4 \pi c p^3 f(p) 
		= \frac{4 \pi c f_i p_i^4}{\rho} \frac{\left( \left( \frac{p_{i+1}}{p_i}\right)^{4-q_i} - 1 \right)}{4 - q_i } \label{eq:e_i_analytic}
\end{align}
Both these equations contain a $\frac{1}{\rho}$ term due to the Lagrangian reference frame in which we discretize the equations, which represents these quantities per unit mass.
To simplify and unify the descriptions for CR electrons and protons we choose to represent the momentum in dimensionless units, meaning $\hat{p} \equiv \frac{p}{m_i c}$. 
To avoid a division by zero for $q_i = 3$ and $q_i = 4$ we introduce a slope softening parameter $\epsilon = 10^{-6}$ and interpolate, as an example for $N_i$
\begin{equation}
    N_i = \frac{4 \pi f_i p_i^3}{\rho } \left( \frac{q_i-3}{\epsilon} + \log\left(\frac{p_{i+1}}{p_i}\right)\left( 1 - \frac{q_i-3}{\epsilon} \right) \right)
\end{equation}
Evolving both CR number and energy is commonly referred to as a \textit{two-moment approach} \citep[see e.g.][for a review on different approaches]{Hanasz2021}.
We use this two-moment approach for both electrons and protons.
For electrons this is required to accurately capture rapid cooling processes.
For protons the second moment is often neglected based on the assumption of a quasi stationary spectrum or a spectrum of constant curvature \citep[][]{Miniati2001}. However, this can lead to numerical instabilities if energy is injected only into one part of the spectrum \citep[see][for a detailed discussion of this problem]{Girichidis2020}.
\subsection{Boundary Conditions}
\label{sec:boundary_conditions}
To decouple the CR component from the non-relativistic gas we need to set the boundary conditions of our distribution function accordingly.
We choose $\hat{p}_\mathrm{min}$ and $\hat{p}_\mathrm{max}$ arbitrarily to best fit the requirements of the simulation and do not neccessarily want to cover the full scale of relativistic energies. This allows us to save on memory and computational cost in cases where we are only interested in the high-energy range of the distribution function.
Therefore, we have to handle the treatment of CRs that move out of this range.
For the presented work we choose open boundaries at the lower end of the distribution function and closed boundaries at the upper end.
The physical motivation for this is that as CRs cool and lose energy they smoothly transition to the non-relativistic thermal background of particles.
For our purposes we assume a gap in the transition between the Maxwell-Boltzmann distribution of particles and the power-law high-energy tail.
By doing so we can treat our implementation as a two component fluid with distinct equations of state with a sharp jump between the two and thus work around the problem of an intermediate state.\\
The upper end of the distribution function is chosen to have a closed boundary.
In order to achieve this we employ a movable upper boundary that also works as a cutoff of the distribution function $p_\mathrm{cut}$.
This parameter needs to be updated at \textit{every} time step.\\
Here the physical motivation is that particles can be further accelerated beyond the arbitrarily chosen initial upper limit of the distribution function. Numerically we avoid an artificial pile-up of energy and particles in the last momentum bin.
In the case of electrons this treatment also leads to a more accurate description of the cooling spectrum. As electrons lose energy, the high momentum end of the distribution function is depopulated. Since we cannot assume that over a time step of our simulation a whole bin is depopulated, we need to be able to represent the distribution function with partially filled bins. To accomplish this we can employ the spectral cut and solve the number- and energy density integrals between the lower bin boundary and the cutoff within the bin.
%
%
\subsection{Time Evolution}
\label{subsec:time_evolution}

We evolve Eq.~\ref{eq:fp-xdiffusion} in time by use of operator splitting. For this we treat the individual terms as independent where necessary and combined where possible. We will give a brief derivation of this process in the following sections.

\subsubsection{Particle Number and Energy Changes}
The key point of evolving the distribution function is to trace the changes in number- and energy-density as a function of time.
For simplicity we follow the approach by \cite{Miniati2001} and show this for adiabatic changes and radiative losses.
The part of Eq.~\ref{eq:fp-xdiffusion} governing those effects is
\begin{equation}
	\frac{D f}{D t} = \frac{1}{3} \frac{\partial u}{\partial x} p \frac{\partial f}{\partial p} + \frac{1}{p^2} \frac{\partial}{\partial p} (p^2 b_l f) \:\: .
	\label{eq:diff_conv_adiab_rad}
\end{equation}
By multiplying both sides of Eq.~\ref{eq:diff_conv_adiab_rad} with $4\pi p^2/\rho$ to be able to associate the l.h.s. with Eq.~\ref{eq:n_i_analytic} and integrating both sides over the bin with index $i$ yields:
\begin{equation}
	\frac{D N_i}{D t} = \frac{1}{\rho} \left\lbrace \left( \frac{1}{3} \frac{\partial u}{\partial x} p + b_l(p) \right) 4 \pi p^2 f(p) \right\rbrace_{p_i}^{p_{i+1}} \:\:.
\end{equation}
As proposed by \cite{Miniati2001} we can integrate this in time and can identify the r.h.s. as the time averaged fluxes over the momentum boundaries.
This gives a number density of bin $i$ after a timestep $\Delta t$
\begin{equation}
	N_i^{t+\Delta t} = N_i^t + \frac{1}{\bar{\rho}} \left(F_{N_{i+1}}^m - F_{N_i}^m \right)
	\label{eq:n_update}
\end{equation}
where $F_{N_{i+1}}^m$ and $F_{N_i}^m$ are the CR number fluxes into and out of the bin respectively and $\bar{\rho}$ denotes the mean density over the timestep.
Similarly, to obtain the energy after a timestep we multiply Eq.~\ref{eq:diff_conv_adiab_rad} with $4\pi c p^3/\rho$ to be able to associate the l.h.s. with Eq.~\ref{eq:e_i_analytic} and perform the same integral as before.
This gives
\begin{align}
	\frac{D E_i}{D t} =& \frac{1}{\rho} \left\lbrace \left( \frac{1}{3} \frac{\partial u}{\partial x} p + b_l(p) \right) 4 \pi c p^3 f(p) \right\rbrace_{p_i}^{p_{i+1}} \\&- \left( \frac{4}{3} \frac{\partial u}{\partial x} E_i + \frac{1}{\rho} \int\limits_{p_i}^{p_{i+1}}  dp \: b_l(p) 4 \pi c p^2 f(p) \right) \:\: .
\end{align}
To simplify this equation we can introduce the quantity $R_i(q_i, p_i)$ for the energy loss integral per bin
\begin{equation}
	R_i(q_i, p_i) = \frac{4 - q_i}{p_{i+1}^{4-q_i} - p_{i}^{4-q_i}} \int\limits_{p_i}^{p_{i+1}}  dp \:\: p^{2 - q_i} \left( \frac{1}{3} \frac{\partial u }{\partial x} + \sum_{l}^{N_\mathrm{losses}} b_l(p) \right) \:\: .
	\label{eq:R_losses}
\end{equation}
Here again $\frac{1}{3} \frac{\partial u }{\partial x}$ denotes the adiabatic changes and $\sum_{l}^{N_{\mathrm{losses}}} b_l(p)$ the individual energy loss processes. In most cases these factors cannot be solved in one step, but need to be split up in individual processes.
As in \citet{Miniati2001} we can express the energy in a bin after the timestep $\Delta t$ as
\begin{equation}
	E_i^{t + \Delta t} \left( 1 + \frac{\Delta t}{2} R_i(q_i, p_i) \right) = E_i^t \left( 1 - \frac{\Delta t}{2} R_i(q_i, p_i) \right) + \frac{1}{\bar{\rho}} \left( F_{E_{i+1}}^m - F_{E_i}^m \right)
	\label{eq:e_update}
\end{equation}
with $\bar{\rho}$ being the mean density over the timestep, $F_{E_{i+1}}^m$ the energy flux into the bin and $F_{E_i}^m$ the energy flux out of the bin.
%
%
\subsubsection{Fluxes Between Momentum Bins}
\label{sec:fluxes}
We can now identify the first term of the integral by parts as the time averaged fluxes over one bin boundary as
\begin{align}
	F_{N_i}^m &= \int\limits_{t}^{t+\Delta t} dt' b_l(p) 4\pi p^2 f(t', p)\vert_{p_i} \\ 
	F_{E_i}^m &= \int\limits_{t}^{t+\Delta t} dt'  b_l(p) 4\pi c p^3 f(t', p)\vert_{p_i}
	\label{eq:flux_of_t}
\end{align}
This is a consequence of our fixed momentum boundaries.
A particle of momentum $p_u$ gains, or looses momentum over a timestep $\Delta t$ and arrives at momentum $p_i$.
Since we use fixed momentum boundaries we cannot move those boundaries to account for this change and instead need to calculate a flux over the boundary into a higher, or lower bin \citep[for an alternative langrangian approach to bin boundaries see][]{Mimica_2009}.
If we consider the definition of our momentum changes
\begin{align}
	\frac{dp}{dt} = b_l(p)
	\label{eq:def_mom_changes}
\end{align}
we can solve this for $dt$ and substitute $dt'$ in Eq.~\ref{eq:flux_of_t}. This gives an equation for the fluxes only dependent on $p$
\begin{align}
	F_{N_i}^m &=  \int\limits_{p_i}^{p_u}  \: dp \: 4 \pi p^2 f^m(p) \label{eq:n_flux_integral}  \\
	F_{E_i}^m &=  \int\limits_{p_i}^{p_u}  \: dp \: 4 \pi c p^3 f^m(p), \label{eq:e_flux_integral} 
\end{align}
with 
\begin{align}
	f^m(p) &= \begin{cases} f_i \left(\frac{p}{p_i}\right)^{-q_i} \quad & \mathrm{if } p_u > p_i \\
							f_{i-1} \left(\frac{p}{p_{i-1}}\right)^{-q_{i-1}} \quad & \mathrm{if } p_u \leq p_i,
			  \end{cases} 
\end{align}
and $p_u$ being the momentum a particle needs to have to yield the momentum $p_i$ after a timestep $\Delta t$.
To solve this integral we use separation of variables in Eq.~\ref{eq:def_mom_changes}
\begin{align}
	\Delta t &= \int\limits_{p_i}^{p_u}  \frac{dp}{b_l(p)} \: .
	\label{eq:flux_integrals}
\end{align}
Identifying $\Delta t$ as the timestep of our simulation and $p_i$ as the momentum bin $i$ we only need to calculate adiabatic changes and radiative losses, respectively to find $p_u$ and with that the lower boundary of the flux integral. 
%
%
\subsubsection{Spectral Cut Update}

As discussed in the section about boundary conditions we chose to keep the upper boundary of our distribution closed and allow no in- or outflux. 
This leads to a right-shift of the cutoff momentum of the distribution in the case of energy gains and a left-shift in the case of energy losses. 
The spectral cutoff needs to be updated at every step of the spectral evolution.
Similar to the fluxes the spectral cutoff can be updated by solving for the integration boundaries in Eq.~\ref{eq:flux_integrals}. Here we assume that the cutoff is the momentum at which the most energetic particles arrive after a timestep $\Delta t$ by setting $p_\mathrm{cut} = p_u$ and solving Eq.~\ref{eq:def_mom_changes} for $p_i$.
%
%
\subsubsection{Slope Update}

As a next step we need to update the slope of each momentum bin. Solving Eqs. \ref{eq:n_i_analytic} and \ref{eq:e_i_analytic} for $f_i$ gives an equation only dependent on $E_i$, $N_i$ and $q_i$.
\begin{align}
	\frac{E_i}{N_i p_{i} c} = \frac{3 - q_i}{4 - q_i} \frac{\left( \frac{p_{i+1}}{p_i} \right)^{4-q_i}-1}{\left( \frac{p_{i+1}}{p_i} \right)^{3-q_i}-1} 
	\label{eq:slope_solver}
\end{align}
Having calculated $E_i$ and $N_i$ in the previous steps we can solve Eq.~\ref{eq:slope_solver} for $q_i$ with any suitable root-finding method.
This is usually done via the Newton-Rhapson method \citep[e.g.][]{Miniati2001, Girichidis2020, Ogrodnik2020}, which shows fast convergence, but requires an initial guess. This guess can either be provided in tabulated form as in \citet{Girichidis2020} and \citet{Ogrodnik2020} or has to be found in a preparation step e.g. by a bracketing method which can prove to be expensive. In this work we find faster convergence using Brent's method at the same accuracy. As this is the most expensive computational step of the scheme we get a substantial performance boost from this choice. We discuss the performance impact briefly in Appendix \ref{app:performance_scaling}.
To further reduce the cost of this step we introduce a finite search range for the root finding with $q \in [-20, 20]$. As bins with a slope $\vert q \vert > 20$ will contribute very little to the overall number- and energy density we accept this artificial error for the benefit of reduced computational cost.
%
%
\subsubsection{Norm Update}

With all other variables updated we can update the normalization of the distribution function. This can in principle be done by solving either of Eq.~\ref{eq:n_i_analytic} or \ref{eq:e_i_analytic} for $f_i$. In practice it is slightly cheaper to solve Eq.~\ref{eq:n_i_analytic} so that the new normalisation of bin $i$ can be computed from
\begin{align}
	f_i = \frac{\rho \: N_i}{4 \pi p_i^3} \frac{3 - q_i}{\left( \frac{p_{i+1}}{p_i} \right)^{3 - q_i} - 1}
	\label{eq:f_i_update}
\end{align}
Ideally one could also solve both Eqs. and construct an interpolation scheme between the two to reduce errors. This could be tested for potential benefits in future work.
%
%
\subsection{Spatial Propagation}
The propagation of CRs in physical space and the physical processes involved have been the matter of quite some debate
\citep[for a recent review on simulations of CR propagation see][]{Hanasz2021}.
As proper discussion of these processes are beyond the scope of this work and propagation is only of minor importance for the system that we are studying in the remainder of this paper (see Appendix \ref{app:diff} for a discussion of the comparison of the timescales involved), we shift the proper description of our diffusion model to follow-up work.\\
However, we adopted a simplified version an isotropic diffusion model to counter numerical noise introduced by the shock finder in simulations where no physical diffusion is required.
This is for example the case in the idealized cluster merger simulation in Section \ref{sec:cluster_mergers} where only initial acceleration is modeled and propagation times are longer than the relevant cooling times of synchrotron bright CR electrons.\\
In the case of simplified diffusion we update the quantity $Q$ of particle $i$ based on the neighboring particles $j$ with
\begin{equation}
    \frac{\mathrm{d} Q_i }{\mathrm{d} t} = \sum_j \frac{m_j}{\rho_i \rho_j} \: \kappa_c \: v_{\mathrm{sig}} \: ( Q_j - Q_i ) \nabla_i W_{ij} 
\end{equation}
where $\kappa_c$ is a constant diffusion coefficient and $v_{\mathrm{sig}}$ is the signal velocity of the CRs, which in the simplest case is equal to the Alfvén velocity.
For the current work we use $\kappa \approx 5 \times 10^{26} \frac{\mathrm{cm}^2}{\mathrm{s}}$ and the Alfvén velocity in the ICM is typically of the order $v_A = \frac{B}{\sqrt{\rho}}\approx 10^3 \frac{\mathrm{km}}{\mathrm{s}}$.
We find that even this simple approach conserves the total energy to a relative error of only $0.2$ per cent over 1 Gyr.

%
\subsection{Adiabatic Changes}
With CRs being confined within the surrounding gas by the CR streaming instability due to their scattering at (self-excited) Alfvén waves \citep[e.g.][]{Kulsrud1969,Wentzel1974, Skilling1975_I, Skilling1975_II, Skilling1975_III} they are dynamically coupled to this gas. As this Alfvén rest frame is compressed, the CRs gain energy based on the PdV  work of the gas. Given that the Alfvén waves have sufficiently high modes this process should be self-similar, so every particle should gain the same amount of energy. 
In the case of a power-law distribution of particles this should contain the power-law shape and only shift to higher energies and momenta respectively.
This leaves the problem of how to handle the lower end of the distribution. In previous works this has been addressed by setting a lower cut \citep[e.g.][]{Winner2019, Ogrodnik2020} or a larger 0th bin as a buffer zone with open lower boundary conditions \citep[e.g.][]{Girichidis2020, Girichidis2022}. 
As discussed in Section \ref{sec:boundary_conditions} we choose to keep an open boundary condition at the lower end of the spectral distribution.
The influx can be achieved by interpolating the lowest momentum boundary to a ``ghost bin'' ($p_\mathrm{g}$) and solving the flux over the lowest boundary
\begin{equation}
	p_\mathrm{g} = p_0 \cdot 10^{-\Delta p}
\end{equation}
where $p_\mathrm{g}$ is the boundary of the ghost bin, $p_0$ is the boundary lowest bin and $\Delta p$ is the bin-width of the spectrum. The normalization of the ghost bin can then be interpolated as
\begin{equation}
	f_g = f_0 \left( \frac{p_0}{p_g} \right)^{q_0}
\end{equation}
where again $f_0$ is the norm and $q_0$ is the slope of the lowest bin.\\
The momentum change due to adiabatic expansion or compression of the surrounding gas can be described by
\begin{equation}
	\left( \frac{\partial p}{\partial t} \right)_\mathrm{adiab.} = - \frac{1}{3} \frac{\partial u}{\partial x} p = - \frac{1}{3} \ln \left( \frac{\rho}{\rho_0} \right) \frac{p}{\Delta t} \:\: .
\end{equation}
Integrating this by parts, as described in Section \ref{sec:fluxes}, and solving the momentum integral for the upper boundary yields
\begin{equation}
	p_u = p_i \left( \frac{\rho_t}{\rho_{t+\Delta t}} \right)^{1/3} \:\: .
\end{equation}
This boundary can then be inserted into the flux integrals in Eqs.~\ref{eq:n_flux_integral} and \ref{eq:e_flux_integral} to compute the number- and energy-density fluxes between momentum bins.

\subsection{Radiative Energy Losses}

For the high momentum end of the CR electron distribution the dominant loss mechanism are inverse-Compton scattering of electrons on CMB photons and synchrotron losses due to the surrounding magnetic field. These loss mechanism both scale with $p^2$ and only depend on the energy density of the background photon field and the magnetic field, respectively. This makes it convenient to combine them into one loss process. The momentum change for a particle due to inverse compton scattering (IC) and synchrotron losses can be written as
\begin{align}
	\left( \frac{dp}{dt} \right)_{\mathrm{Synch+IC}} = \frac{4}{3} \frac{\sigma_T}{m_e^2 c^2} \: \left(U_{\mathrm{IC}} + U_\mathrm{B} \right) \: p^2 = \beta \: p^2
	\label{eq:ic_synch_e}
\end{align}
where we introduced $\beta = \frac{4}{3} \frac{\sigma_T}{m_e^2 c^2} \: \left(U_{\mathrm{IC}} + U_\mathrm{B} \right)$ for convenience.
Following the steps in Section \ref{subsec:time_evolution} we can solve this for the upper integration boundary as
\begin{align}
    p_u = \frac{p_i}{ 1 - \beta \: p_i \: \Delta t},
\end{align}
and update the spectral cut as
\begin{align}
	p_{\mathrm{cut},t+\Delta t} = \frac{p_u}{1 + \beta p_{\mathrm{cut},t} \Delta t} \:\: .
	\label{eq:ic_synch_cut}
\end{align}
With that we can solve the flux integrals (Eq.~\ref{eq:e_flux_integral} and \ref{eq:n_flux_integral}) and the number density update (Eq.~\ref{eq:n_i_analytic}). To evolve the energy density we also need to solve Eq.~\ref{eq:R_losses} per bin as
\begin{align}
    R_i(q_i, p_i) =  \beta \: \frac{4 - q_i}{p_{i+1}^{4-q_i} - p_{i}^{4-q_i}} \:\: \frac{p_{i+1}^{5-q_i} - p_{i}^{5-q_i}}{5 - q_i}
\end{align}

\subsection{Source Terms for CRs}

In our model we account for the sources of CRs in our simulations based on structure formation shocks and SNe.
For the present work only injection at shocks is of relevance, we will therefore introduce further injection models in future work where it is applicable.
We will describe the injected energy and spectra in the following subsections.
To identify shocks in our simulations we use the on-the-fly shockfinder introduced in \citet{Beck2016b}.

\subsubsection{Shock Acceleration}
\label{sec:shock_acceleration}

To bridge the gap between the small-scale physics of DSA and large-scale shocks in the ICM we take the result from PIC simulations and include them in a subgrid description. We quantify this as different models of acceleration efficiencies $\eta$ that depend on the sonic mach number $\mathcal{M_\mathrm{s}}$, the ratio between upstream thermal and CR pressure $X_\mathrm{cr} \equiv \frac{P_\mathrm{cr,u}}{P_\mathrm{th,u}}$ and the angle between magnetic field and shock normal $\theta_\mathrm{B}$.
Generally the energy injected into a CR population behind a shock can be written as \citep[e.g.][]{Kang2007}
\begin{equation}
	E_{\mathrm{CR},2} = \: \eta(\mathcal{M}_\mathrm{s}, X_\mathrm{cr}) \: \eta(\theta_\mathrm{B}) \: E_{\mathrm{sh}}
	\label{eq:postshock_e_corrected}
\end{equation}
Here $E_{\mathrm{sh}}$ is the energy dissipated at the shock and the two efficiency functions $\eta(\mathcal{M}_\mathrm{s}, X_\mathrm{cr})$ and $\eta(\theta_\mathrm{B})$ describe which fraction of that shock energy is injected due to the strength of the shock $\eta(M,X_{cr})$ and the geometry between magnetic field vector and shock normal $\eta(\theta_B)$.
We use two different methods to obtain the shock energy. One is via the on-the-fly shock finder
\begin{equation}
    E_{\mathrm{sh}} = \frac{1}{2} u_{\mathrm{sh}}^3 \: \frac{\Delta t}{2h_{i}}
\end{equation}
where $u_{\mathrm{sh}}$ is the shock speed obtained from the shock finder, $\Delta t$ is the time step and $h_{i}$ is the hydrodynamic smoothing length.
This denotes the shock energy per timestep, normalized to conserve the total energy the shock dissipates as it runs through the region broadened by the SPH kernel. 
This works well in idealized simulations such as shock tubes, but is prone to numerical noise and limitations from the shock finder in resolution limited cases.
As the shock is numerically broadened and detected slightly in front of the actual shock front, the injection also happens in the pre-shock region.
This in turn leads to a precursor wave of CR gas which can lead to a runaway effect for high CR injection efficiencies.
This is often remedied by saving the shock energy and injecting it after a delay time into the post-shock region \citep[e.g.][]{Pfrommer2006, Pfrommer2016, Dubois2019} which in turn reduces the temporal resolution of the injection mechanism.\\
As an alternative method we compute the shock energy from the entropy change per timestep.
\begin{equation}
    E_\mathrm{sh} = \frac{ \Delta S  }{ ( \gamma - 1) \rho^{ \gamma - 1} } \: \Delta t
\end{equation}
This has the advantage of being numerically self-consistent as it represents the actual energy dissipated as computed in the hydro solver, instead of the shock finder.
The downside is again that the shock finder detects the shock slightly in front of the actual shock front, which leads to the injection process not capturing the whole shock time.
For Mach number dependent acceleration efficiencies this has the additional disadvantage of only capturing the decaying flank of the broadened shock, which leads to an additional under-prediction of the acceleration efficiency.
Nonetheless, both these effects can be countered by tuning on shock tubes, which makes the entropy injection method more stable than the shock speed injection method in our tests.

\subsubsection{Mach Number Dependent Efficiency Models}

\begin{table}
    \centering
    \caption{From left to right we report the name, the ratio between thermal and CR pressure to use for the interpolation between acceleration and re-acceleration, the values for the parameters of Eq.~\ref{eq:eta_kang07_acc} and the critical sonic Mach number from the different efficiency models.}
    \begin{tabular}{|l|c|c|c|c|c|c|c|}
    \toprule
    Model & $X_\mathrm{cr,0} $& $b_0$ & $b_1$ & $b_2$ & $b_3$ & $b_4$ & $\mathcal{M}_\mathrm{s, crit}$\\
    \midrule
    \midrule
        KR07  & 0 & 5.46 & -9.78 & 4.17 & -0.33 & 0.57 & 1 \\
        KR07$r$ & 0.3 & 0.24 & -1.56 & 2.8 & 0.51 & 0.56 & 1 \\
        KR13  & 0 & -2.87 & 9.67 & -8.88 & 1.94 & 0.18 & 2\\
        KR13$r$ & 0.05 & -0.72 & 2.73 & -3.29 & 1.34 & 0.19 & 2 \\
        Ryu19  & 0 & -1.53 & 2.40 & -1.25 & 0.22 & 0.03 & 2.25\\
        Ryu19$r$ & 0.05 & -0.72 & 2.73 & -3.29 & 1.34 & 0.19 & 2.25\\
    \bottomrule
    \end{tabular}
    \label{tab:eff_model_values}
\end{table}

\begin{figure}
	\centering
	\includegraphics[width=\columnwidth]{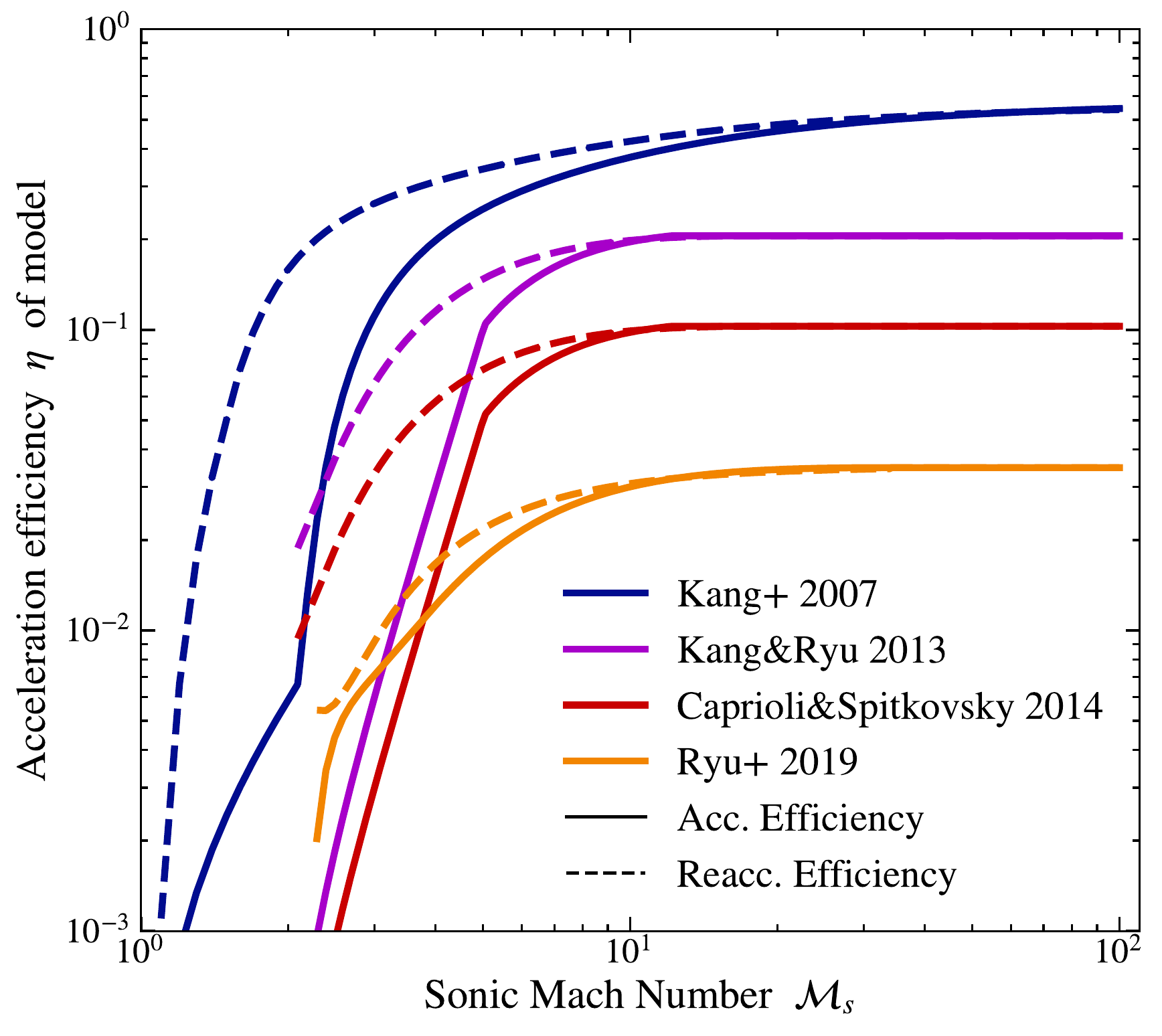}
	\caption{Mach number dependent efficiency models present in our model. The x-axis shows the Mach number and the y-axis the fraction of shock energy that is being injected into the CR component. Solid lines indicate initial acceleration, dashed lines indicate re-acceleration.}
    \label{fig:eff_models}
\end{figure}

For this work we implemented four different Mach number dependent efficiency models introduced by \citet{Kang2007, Kang2013, Caprioli2014} and \citet{Ryu2019} for physical systems and the constant injection efficiency used by \citet{Pfrommer2016} and \citet{Pais2018} for test problems.
To the best of our knowledge only \citet{Kang2007} provide a fitting function to their data with
\begin{align}
  \eta(\mathcal{M}_s) = \begin{cases}
  				1.96\cdot 10^{-3} (\mathcal{M}_\mathrm{s}^2 - 1)	\: & \mathrm{for} \: \mathcal{M}_\mathrm{s} \leq 2 \\
  			    \sum_{n=0}^4 b_n \frac{(\mathcal{M}_\mathrm{s}-1)^n}{\mathcal{M}_\mathrm{s}^4} &  \mathrm{for} \: \mathcal{M}_\mathrm{s} > 2
  			\end{cases}
  \label{eq:eta_kang07_acc}
\end{align}
for initial acceleration and only the equation for $\mathcal{M}_\mathrm{s} > 2$ for re-acceleration. 
We find that these equations also provide a good basis for fitting to the remaining Mach number dependent acceleration functions.
Table~\ref{tab:eff_model_values} gives a reference for all the values of $b_n$ used in our description of the different acceleration models\footnote{We also provide a public version of the DSA models at \url{https://github.com/LudwigBoess/DSAModels.jl}}, which we obtained from fitting their published data.
The model by \citet{Kang2013} for initial acceleration (KR13) can be well described with
\begin{align}
  \eta(\mathcal{M}_s) = \begin{cases}
  				c + d \mathcal{M}_\mathrm{s}^e 	\: & \mathrm{for} \: 2 \leq \mathcal{M}_\mathrm{s} \leq 5 \\
  			    \sum_{n=0}^4 b_n \frac{(\mathcal{M}_\mathrm{s}-1)^n}{\mathcal{M}_\mathrm{s}^4} &  \mathrm{for} \: 5 < \mathcal{M}_\mathrm{s} \leq 15\\
  			    0.211 &  \mathrm{for} \: \mathcal{M}_\mathrm{s} > 15
  			\end{cases}
  \label{eq:eta_kang13_acc}
\end{align}
with $c = -5.95 \cdot 10^{-4}$, $d = 1.88\cdot 10^{-5}$ and $e = 5.334$. We also introduced a saturation value of $\eta_\mathrm{max} = 0.211$ for $\mathcal{M}_\mathrm{s} > 15$.
Their re-acceleration model (KR13$r$) is well fit by the $\mathcal{M}_\mathrm{s} > 2$ form of Eq.~\ref{eq:eta_kang07_acc} with the parameters given in Tab.~\ref{tab:eff_model_values}. Both these models assume that only shocks with $\mathcal{M}_\mathrm{s} > 2$ can efficiently accelerate particles.
We do not explicitly list the values of \citet{Caprioli2014} (CS14), as we take the same approach as \citet{Vazza2016} and assume that the efficiency is half that of KR13 and KR13$r$.
\citet{Ryu2019} performed their study only for sonic Mach numbers relevant for intra-cluster shocks in the range of $2.25 \leq \mathcal{M}_\mathrm{s} \leq 5$ where only supercritical shocks can accelerate CRs \citep[motivated by the findings of][]{Ha2018}. 
To be able to account for higher Mach number shocks we interpolate their data up to higher Mach numbers assuming a similar functional form as the re-acceleration model KR13$r$ with a maximum efficiency of $\eta_\mathrm{max} = 0.035$.
We find that both initial acceleration (Ryu19) and re-acceleration (Ryu19$r$) are well described by using the values for $b_n$ listed in Tab.~\ref{tab:eff_model_values}.\\
We interpolate between acceleration and re-accleration models based on the CR to thermal pressure ratio present in the SPH particle.
As a basis for this we use the seeded CR population of the underlying models $X_\mathrm{cr,0}$ (given in Tab.~\ref{tab:eff_model_values}) and interpolate linearly between the two models according to $X_\mathrm{cr}$ contained in the particle.
Fig \ref{fig:eff_models} gives a visualisation of all models.
\subsubsection{Magnetic Field Geometry Dependent Efficiency Models}
As noted above, the magnetic field morphology at the shock plays a vital role in the triggering of instabilities and with that the acceleration efficiency. Unfortunately, these instabilities are significantly below the resolution limits of current large scale hydrodynamical simulations. Most recent work treats these processes as sub-grid models and use a statistical approach to give an additional efficiency parameter \citep[e.g.][]{Vazza2016}, or just allow CR injection in a specific angle range and switch acceleration on and off \citep[e.g][]{Banfi2020}.\\
In this work we take the same approach as \citet{Pais2018, Dubois2019} and introduce an additional factor $\eta(\theta_B)$ in our total acceleration efficiency. This parameter was obtained by \citet{Pais2018}, who use the values by \citet{Caprioli2014} to fit a functional form to their data as
\begin{equation}
	\eta(\theta_\mathrm{B}) \approx \frac{1}{2} \left[ \tanh \left( \frac{\theta_{\mathrm{crit}} - \theta_\mathrm{B}}{\delta} \right) + 1 \right]
	\label{eq:b-angle-dependence}
\end{equation}
with $\delta = \pi / 18 $ and $\theta_{\mathrm{crit}} = \left( \pi/4 ; \pi/3 \right)$. $\theta_{\mathrm{crit}} = \pi/4$ corresponds to a shock without and $\theta_{\mathrm{crit}} = \pi/3$ to a shock with pre-existing CR component \citep[from][respectively]{Caprioli2014, Caprioli2018}. These efficiencies were modeled for ions, for which DSA should be most effective at quasi-parallel shocks. For electrons quasi-perpendicular shocks should be the main driver of acceleration, as outlined above. We therefore take the simple approach of shifting the efficiency model by $90^{\circ}$ for electrons for the purpose of this work.\\
\subsubsection{Injection Momentum}
Since we arbitrarily set our lower momentum boundary we need to pay attention to the connection between thermal and non-thermal component. With a fixed lower boundary the spectral connection between the Maxwell-Boltzmann distributed thermal gas and the non-thermal power-law tail is not necessarily represented. We remedy this by again using the results from PIC simulations \citep[e.g.][]{Caprioli2014, Ryu2019} who find the momentum $p_{\mathrm{inj}}$ at which the MBD transitions to a power-law to be a multiple $\chi$ of the momentum of the thermal protons downstream of the shock
\begin{equation}
    p_{\mathrm{inj}} = \chi p_\mathrm{th} = \chi \sqrt{2 m_\mathrm{p} k_\mathrm{B} T_2}
    \label{eq:p_inj}
\end{equation}
where $\chi$ is a free parameter found in the simulations and $T_2$ denotes the gas temperature downstream of the shock. We employ $\chi = 3.5$ and assume that electrons are injected at the same dimensionless momentum as protons.

\subsubsection{Proton to Electron Injection Ratio}

The total energy budget provided by the shock acceleration needs to be distributed over electrons and protons, following some energy ratio. Unfortunately this ratio is poorly constraint with $\frac{f_e}{f_p} \equiv K_{e,p} \sim 0.01 - 0.025$ \citep[e.g.][]{Beck2015}.
As an alternative for the current work we can calculate the electron to proton ratio as found in the semi-analytic approach by \citet{Kang2020} 
\begin{equation}
    K_{e,p} = \left( \frac{m_e}{m_p} \right)^{(q_\mathrm{inj}-3)/2}
    \label{eq:Kep}
\end{equation}
For a typical injection slope of $q_\mathrm{inj} \approx 4-5$ this leads to $K_{e,p} \sim 10^{-2} - 10^{-3}$ \citep[see][for an analogous approach]{Inchingolo2022}.

\subsubsection{Spectral Slope}

In the classic picture of particle acceleration via DSA the acceleration is a self-similar process which converges to a power-law distribution of the particles, in general agreement with observations. A caviat of the standard DSA model \citep[as pointed out by e.g. the review of][]{Drury1983} is that it is based on a purely hydrodynamical shock, while the scattering processes clearly require magnetic fields and with that a magneto-hydrodynamical (MHD) treatment, as outlined above.
A recent set of PIC simulations by \citet{Caprioli2020} \citep[followed up by further investigation by][]{Diesing2021} showed that the shock develops a magnetosonic post-cursor wave that can scatter a large fraction of high-energy CRs out of the acceleration zone.
This leads to a steepening of the spectrum which they parameterize with
\begin{align}
    q = \frac{3r}{r - 1 - \alpha}; \:\: \alpha \equiv \frac{v_{A,2}}{u_2}
    \label{eq:new_dsa_slope}
\end{align}
where $v_{A,2} = B_2 / \sqrt{4 \pi \rho_2}$ and $u_2$ are downstream Alvfén speed and gas velocity, respectively.
They refer to this new description as non-linear diffusive shock acceleration (NLDSA).
It follows trivially that Eq.~\ref{eq:new_dsa_slope} reduces to the standard DSA slope for a non-MHD shock.
We added the computation of $\alpha$ to our on-the-fly shock finder to optionally account for this process.

\subsection{Injection into the Model}

From the source term we obtain three parameters: $E_\mathrm{inj}^\mathrm{CR}$ as the energy to be injected, $p_\mathrm{inj}$ as the momentum at which the injected power-law starts and $q_\mathrm{inj}$, the slope of this power-law. We can then insert these parameters into Eq.~\ref{eq:e_i_analytic} and solve for the normalisation of the distribution function at the injection momentum
\begin{equation}
	f_{\mathrm{inj}} = \frac{E_\mathrm{inj}^\mathrm{CR} ( 4 - q_{\mathrm{inj}} ) }{\frac{4 \pi c \hat{p}_{\mathrm{inj}}^4}{\rho} \left( \left( \frac{\hat{p}_{\mathrm{max}}}{\hat{p}_{\mathrm{inj}}} \right)^{4-q_{\mathrm{inj}}} - 1 \right)} \:\: 
\end{equation}
where $\hat{p}_{\mathrm{max}}$ is the (arbitrarily chosen) upper boundary of the distribution function. This is typically $\hat{p}_{\mathrm{max}} \sim 10^5 - 10^6$. For strongly magnetized shocks this strict power-law injection is typically softened by a exponential cutoff for high momenta in the electron population. In weakly magnetized ICM shocks this can be neglected \citep[see the discussion in][]{Kang2020}. 
The other normalizations can then be interpolated from the power-law shape by using Eq.~\ref{eq:piecewise_powerlaw}.
With the normalisation $f_i$ and slope $q_{\mathrm{inj}}$ of every bin calculated we can inject CR number and energy per bin by solving Eq.~\ref{eq:n_i_analytic} and Eq.~\ref{eq:e_i_analytic} respectively. The spectral cutoff of the distribution is either reset to $p_{\mathrm{max}}$ if it was below that before the injection or kept as is, if it was above $p_{\mathrm{max}}$. To preserve the total energy we subtract the energy injected into the CR component by the shock from the entropy change of the gas component.
Once the energy and CR number of every bin is updated we update the total distribution function by first solving the slope of the individual bins with Eq.~\ref{eq:slope_solver} and then recalculating the normalisation $f_i$ using Eq.~\ref{eq:f_i_update}. 

\subsection{Coupling to the Simulation}

We implemented \textsc{CRESCENDO} into \textsc{OpenGadget3}, a cosmological Tree-SPH code based on \textsc{Gadget2} \citep[][]{Springel2005}.
Due to the lagrangian nature of SPH the update of the hydrodynamical quantities is driven by the total pressure.
To this end we add the CR pressure to the thermal pressure of the particles and use this to update the lagrangian.
Having updated the spectral distribution due to the previously described effects we can now compute the comoving CR pressure component by integrating over the spectrum

\begin{align}
    P_{\mathrm{CR},c} &= \frac{4 \pi}{3} \: a^{4} \: \int\limits_{p_{\mathrm{min}}}^{p_{\mathrm{cut}}} dp \: p^2 T(p) f(p) \\
    & \approx \frac{1}{3} \: a^{4} \: \int\limits_{p_{\mathrm{min}}}^{p_{\mathrm{cut}}} dp \: 4\pi c p^3 f(p) 
    \label{eq:pressure_integral}
\end{align}
where the r.h.s. of Eq.~\ref{eq:pressure_integral} can readily be identified as an energy integral over the whole distribution function. Since we solve the update of the distribution function in physical space for cosmological simulations we introduce the conversion from physical to comoving frame as in \citet{Pfrommer2016} at this point, where $a$ is the cosmological scale factor.
Here we again used the approximation of purely relativistic particles.
This pressure component is then added to the total pressure, which goes into the hydrodynamic acceleration of the SPH particles.
We note that this only provides a lower limit to the total CR pressure, due to the simplification $T(p) \approx pc$.
As we are mainly interested in the high-energy emission of electrons and the observational constraints on CR proton pressure are quite strict, we accept this limitation for the current work.

\subsection{Timestep Constraint}

Similar to \citet{Miniati2001, Yang2017, Ogrodnik2020} we find that the common approach to limit the timestep within the solver so that one bin is not fully depleted within one timestep is not sufficient in the case of fast cooling electrons.
Like the previous authors we therefore employ 
\begin{equation}
    \Delta t_\mathrm{max} \leq 0.1 \tau_{i,\mathrm{cool}}
    \label{eq:dt_limit}
\end{equation}
with $\tau_{i,\mathrm{cool}}$ being the cooling time of each energy loss process.
To avoid computational overhead wherever possible we sub-cycle the solver and update the distribution function at the end of the simulation timestep.

\section{Tests of the CR Model}
\label{sec:tests}

In this section we will outline a number of tests to compare the performance of the model to analytic solutions, where available and test its numerical stability. We will present the tests in the same order as the description of the individual components of the model.

\subsection{Adiabatic Changes}

We test the quality of the adiabatic changes as implemented in our model based on its capability of keeping the spectral slope, as well as its ability to conserve energy throughout every completed model cycle. For completeness, we use two versions of the model, a stand alone version for testing as well as the direct implementation of that model into our code \textsc{OpenGadget3}.
\begin{figure}
	\centering
	\includegraphics[width=\columnwidth]{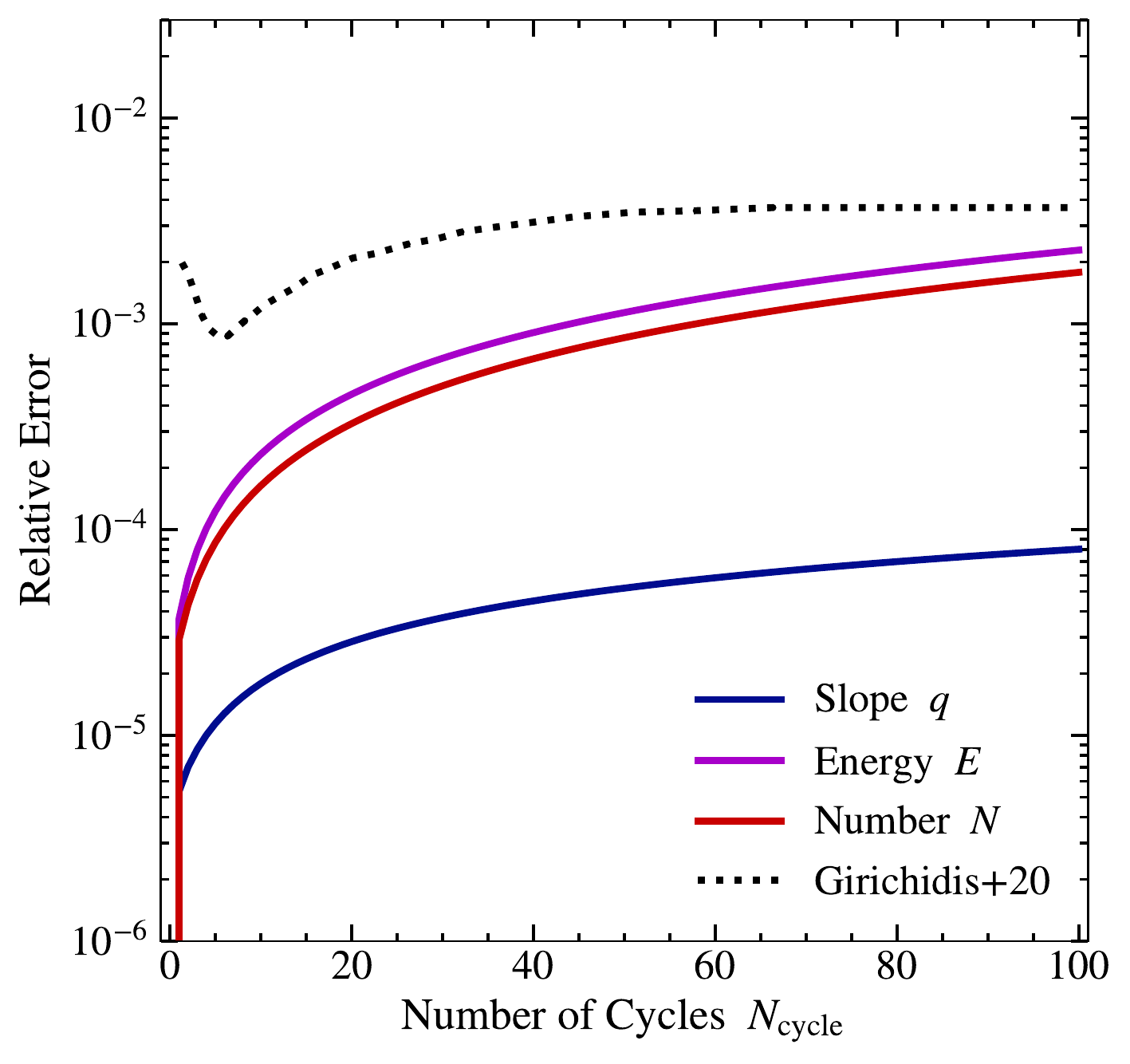}
	\caption{Relative errors for CR energy and number, as well as reconstructed slope in the sine-wave test for a single SPH particle. We represent the spectrum with 2 bins / decade to test a worst-case scenario. The dotted line indicates the upper limit of the error from \citet{Girichidis2020}, Fig.~9 for their piece-wise power-law implementation. We find similar accuracy, but note that while \citet{Girichidis2020} find a stable solution after a number of circles, our errors keep increasing over the course of the test.}
    \label{fig:sinus_standalone}
\end{figure}
For comparison with other implementations we performed the same test as \citet{Girichidis2020} and modeled a sinoidal density wave moving through a single SPH particle. For this we set up a single power-law spectrum with a slope of $q = 4.5$ over six orders of magnitude in momentum. We then set a time-dependent density field as
\begin{equation}
    \rho(t) = \rho_0 \left( 1 + \frac{\sin (2\pi t)}{2} \right)
\end{equation}
and evolve the spectrum for 100 cycles. In addition we run the test with two spectral resolutions, 12 bins and 192 bins or 2 bins/dex and 32 bins/dex, respectively. The result of the $L_1$ error for CR energy / number density and reconstructed slope after every cycle is shown in Fig.~\ref{fig:sinus_standalone}. We find stable behaviour and a comparable accuracy to the implementation by \citet{Girichidis2020}, with the caviat that we find an increasing error after every cycle, while their model appears to stabilize after a number of cycles.
Further investigation shows that this stems from our ghost-bin interpolation. As a small error in the slope reconstruction of the 0-th bin also affects the ghost-bin.
Since the 0-th bin by design contains the most CR energy/number this error can become problematic. We can counter this in future work by either applying a closed lower boundary in simulations where only the upper part of the distribution function is relevant, e.g. in simulations of cosmological structure formation, or by adding low-momentum energy loss processes in simulations of galaxy formation.
For the purpose of this work we accept this behaviour as is, since the overall error is very small.
We only show the result for 12 bins in Fig~\ref{fig:sinus_standalone}, as we find no significance difference in the CR energy and number errors, as is expected due to the nature of the test problem.
Since we set up a single power-law spectrum and adiabatic changes should not change the shape of the spectrum the resolution should not be relevant.
However, in principle more bins have the potential of more numerical inaccuracies, so we find this consistent behaviour to be reassuring.\\
To test the model within \textsc{OpenGadget3} we set up 3D fully hydrodynamic test case of a decaying sine-wave in Appendix \ref{appendix:sinewave}.
There we find excellent numerical stability in a more realistic scenario.
This gives us confidence that the model will behave as expected in production runs.

\subsection{Radiative Cooling}
\label{sec:rad_cooling_test}

To test our model under radiative cooling we set up a small box of SPH particles and switched off all contributions to the spectral evolution except for radiative cooling due to IC scattering of electrons on CMB photons at $z=0$. We initialized the particle spectra as a single power-law with slopes $q_0 = -3.5$ and $q_0 = -6$ in the range $\hat{p} \in [1, 10^6]$ represented by 192 bins, or 32 bins/dex. We evolve the simulation until the cooling time of electrons with momentum $\hat{p} = 10^4$ is reached.
\subsubsection{Accuracy}
\begin{figure*}
	\includegraphics[width=\fullwidth]{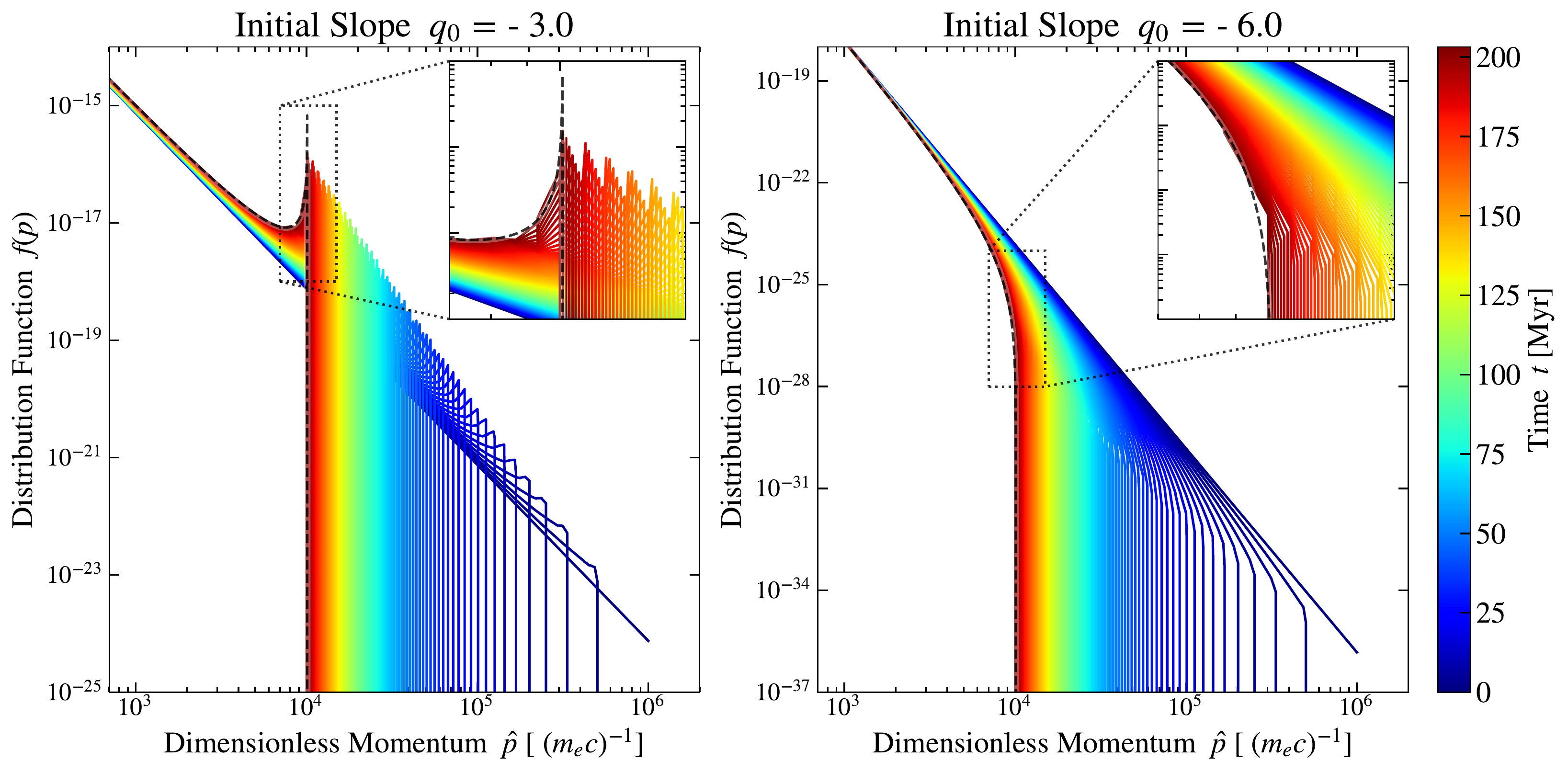}
    \caption{Test particle spectra with 32 bins per order of magnitude in momentum under constant inverse Compton scattering off CMB photons at $z=0$.
    \textit{Left}: Cooling for a spectrum with an initial slope of $q_0 = -3.0$. \textit{Right}: Cooling for a spectrum with an initial slope of $q_0 = -6$. We only show the upper half of the distribution function, as IC losses are only relevant in the high-momentum end. 
    The simulation was run until the cooling time for particles of momentum $\hat{p}_{\mathrm{cool}} = 10^4$ was reached. Colors indicate the spectra at the respective time. The dash line indicates the analytic solution at the final time. We find excellent agreement with the analytic solution and note that the agreement is only limited by spectral resolution.}
    \label{fig:ic_spectra}
\end{figure*}
For testing the accuracy of our radiative cooling implementation we follow \citet{Kardashev1962} who provides an analytic solution for an initial power-law spectrum experiencing radiative cooling from synchrotron radiation and inverse compton scattering. This can be written in terms of the distribution function $f(p)$ as in \citet{Ogrodnik2020}
\begin{align}
  f(p, q, t) = \begin{cases}
  				f(p, t_0) \left( 1 - \beta t p \right)^{q-4}	\: & \mathrm{for } p < \frac{1}{\beta t} \\
  			    0 &  \mathrm{for } p > \frac{1}{\beta t}
  			\end{cases}
  \label{eq:f_kardashev}
\end{align}
where $\beta = \frac{4}{3} \frac{\sigma_T}{m_\mathrm{e}^2 c^2} \: \left(U_{\mathrm{IC}} + U_\mathrm{B} \right)$ as in Eq.~\ref{eq:ic_synch_e}.
This solution indicates a difference in spectral shape for spectra with $q < 4$ and $q > 4$. For $q < 4$ the high-momentum end on the spectrum is so densely populated that cooling particles pile up in lower momentum bins and lead to a flattening and even increase of the spectrum, while for $q > 4$ the high momentum electrons cool off fast enough to lead to a simple steepening of the spectrum. It also predicts a sharp cutoff of the distribution function at $p = \frac{1}{\beta t}$.
The result of this test can be seen in Fig.~\ref{fig:ic_spectra} where we only show the relevant upper half of the spectra. We can see the expected upturn of the spectrum for $q = -3$ and a steepening of the spectrum for $q_0 = -6$ and find very good agreement with the analytic solution (dashed) that is only limited by the spectral resolution of the model.

\subsubsection{Convergence}
\begin{figure}
	\centering
	\includegraphics[width=\columnwidth]{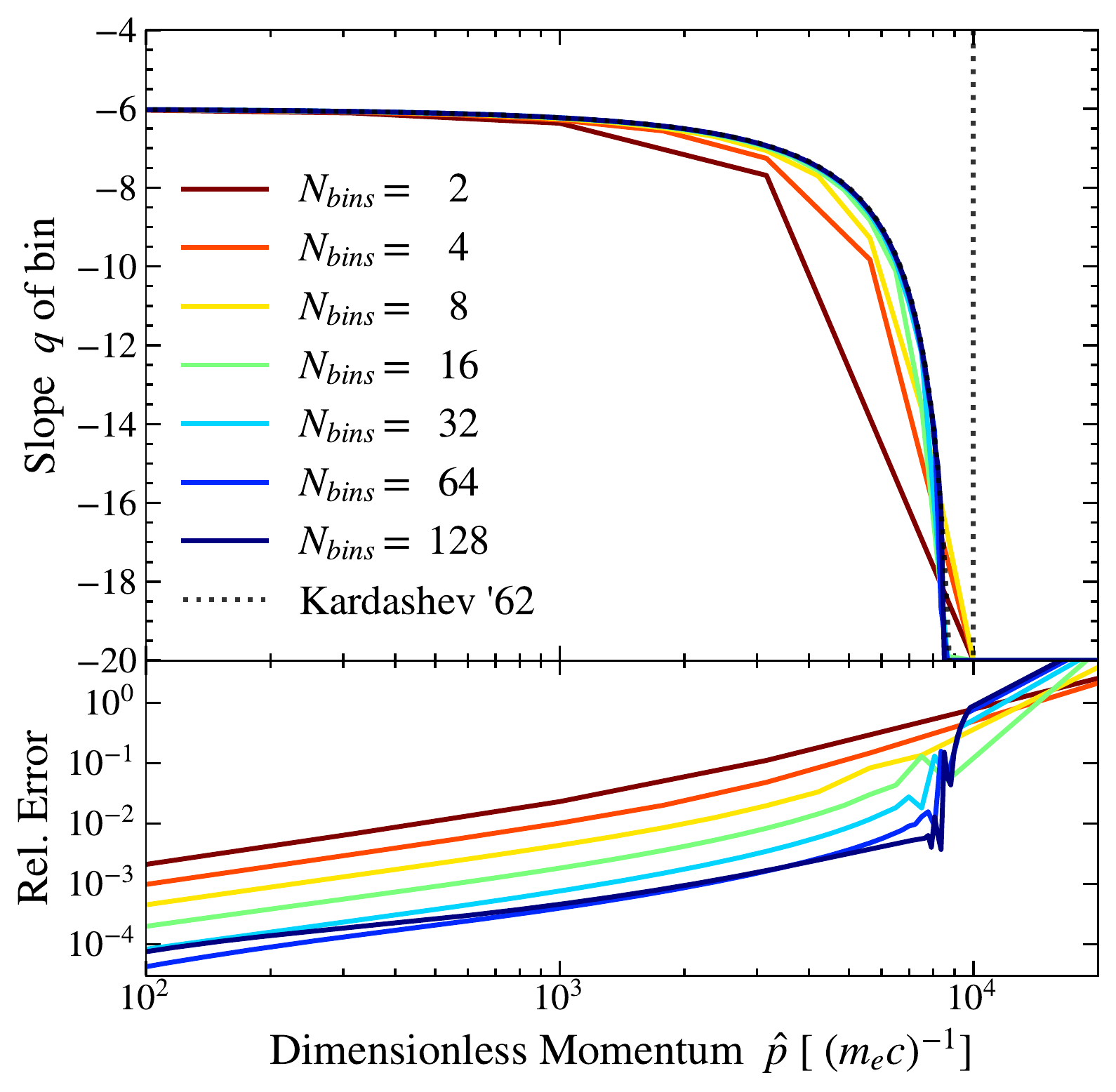}
	\caption{Results of the convergence tests for high-momentum radiative cooling.
	The top panel shows the analytic (dotted) and simulated (solid) slopes for each momentum bin. The dotted vertical line indicates the position of the spectral cutoff.
	The bottom panel shows the corresponding relative error.
	Colors refer to the different number of CR bins per order of magnitude in momentum. We find good convergence to the cooling solution for a spectral resolution of 8 bins per order of magnitude and above.}
    \label{fig:kardashev}
\end{figure}
In order to study the convergence of our model under different spectral resolutions we can rewrite Eq.~\ref{eq:f_kardashev} to represent the spectral slope per bin as a function of time
\begin{equation}
	q(p, t) = - \frac{\partial \ln f}{\partial \ln p} = q_0 + ( q_0 - 4 ) \frac{tp/(\tau_{p_n} p_n)}{1 - tp/(\tau_{p_n} p_n)}
	\label{eq:q_kardashev}
\end{equation}
where $\tau_{p_n}$ is the cooling time for the radiative loss mechanisms.
We repeat the previously described test with different spectral resolutions between 12 bins (2/dex) to 768 bins (128/dex). The results are shown in Fig.~\ref{fig:kardashev}. We find a good convergence trend, with 24 bins (or 4/dex) being the minimum number of bins we consider acceptable to model CR electron cooling due to synchrotron and IC losses.
We note that the discrepancy of the higher resolution models below $\hat{p} = 10^4$ stems from our limit on the slope per bin. As noted above, we ran the simulation until the cooling time of $\hat{p} = 10^4$ is reached and our spectral cutoff also reached that value to very high accuracy. The actual spectrum however should steepen below $q = -20$ and connect to $p_\mathrm{cut}$ at $q \rightarrow -\infty$. This would increase our computing time significantly due to the root finding step, as previously discussed. The bins are therefore artificially set to $q = -20$. We performed the same test for initial slopes of $\vert q_0 \vert < 4$ and found identical convergence behaviour.

\subsection{Shock Injection}

To test our model against an analytic solution we extended the analytic solution derived by \citet{Pfrommer2006} to account for Mach number and magnetic field geometry dependent acceleration efficiencies.
We solve the Riemann problem to first order, higher order solutions would require multiple iterative solution steps for the high efficiency models. As the inclusion of a CR fluid with considerable contribution to the total post-shock energy density slows down the shock \citep[see e.g.][]{Pfrommer2006, Pfrommer2016, Dubois2019} this leads to a lower Mach number and with that a lower acceleration efficiency, which again results in smaller CR component in the post-shock region and a higher Mach number in the next iteration of the solution.
As the more recent acceleration models point to efficiencies below $10$ per cent, this effect becomes considerably smaller than the uncertainty of the models themselves.\\
We list the parameters for all shock tubes used in this section in Table \ref{tab:shocktubes}.

\subsubsection{Mach Number Dependent Efficiency Models}

\begin{figure}
    \centering
	\includegraphics[width=\columnwidth]{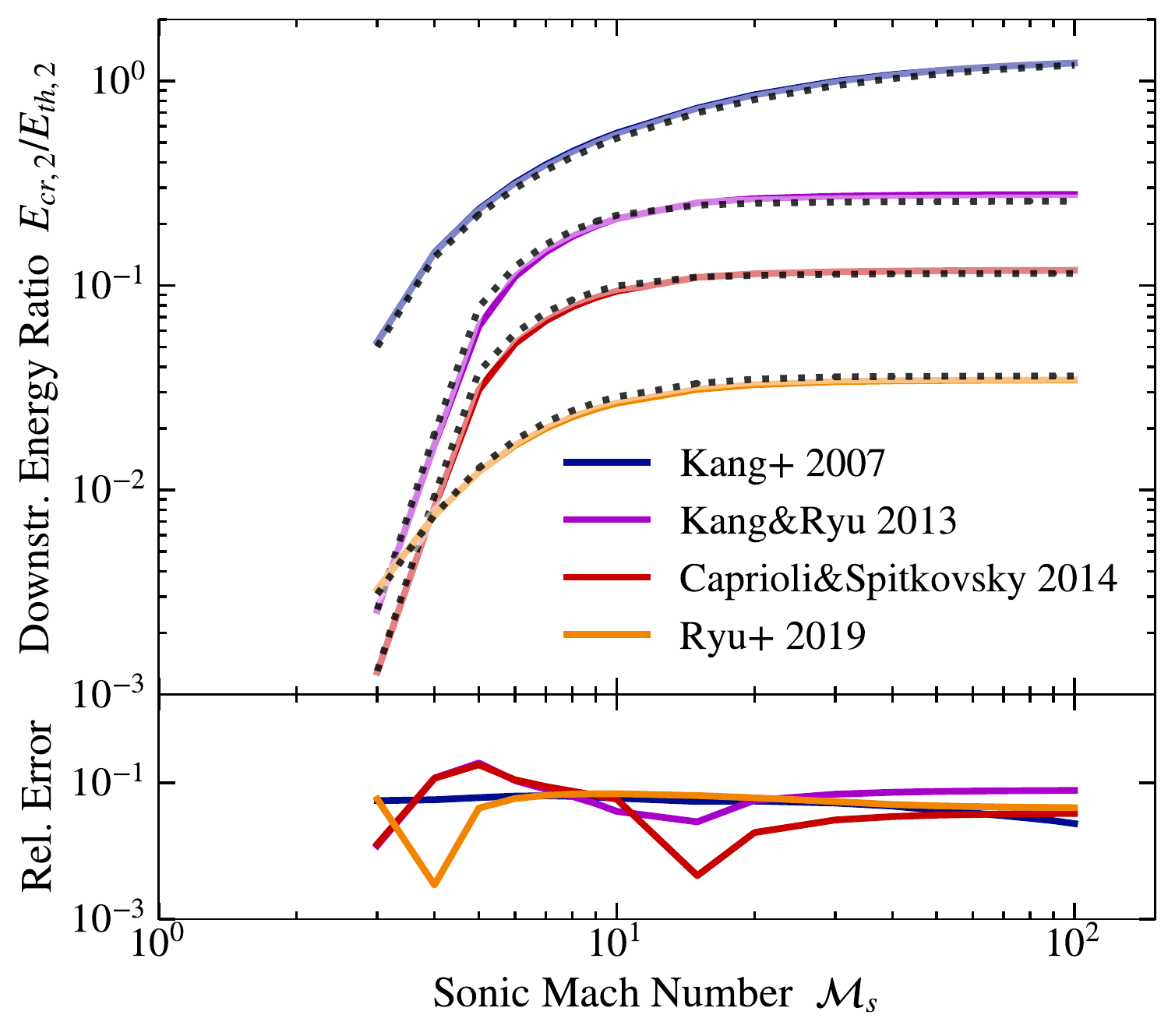}
    \caption{We show the results of the injection tests. The upper panel displays the ratio between downstream CR and thermal energies for a shock with sonic mach number $\mathcal{M}_s$ and the lower panel displays the relative error to the analytic solution. Solid lines indicate the injection method based on entropy change and dotted lines show the analytic solution. The colors refer to the same efficiency models as shown in Fig.~\ref{fig:eff_models}. In general we find excellent agreement with the analytic solution for the entropy changed based injection model.}
    \label{fig:etaM_shocks}
\end{figure}

As a test for the accuracy of our Mach number dependent efficiency $\eta(M)$ we set up a series of Sod shock tubes \citep[following][]{Sod1978}. We used the canonical density jump of $ \rho_L / \rho_R = 8$, kept the left-sided temperature fixed and varied the right-side temperature to obtain resulting shocks with Mach numbers in the range $\mathcal{M}_s \in [3, 100]$.
We ran these shock tubes with all efficiency models shown in Fig.~\ref{fig:eff_models} and with only the proton component switched on. Fig.~\ref{fig:etaM_shocks} shows the result of these tests.
The entropy dependent acceleration method captures the analytic solution quite accurately, with a relative error of 10\% per cent and below.
This is especially evident in the relevant low Mach number regime.
The excellent agreement over all efficiency models together with the little work required to implement them gives us the chance to test upcoming efficiency models in the context of cosmological simulations, as these models become available.
\subsubsection{Magnetic Field Angle Dependent Efficiency Models}
\begin{figure*}
	\includegraphics[width=\fullwidth]{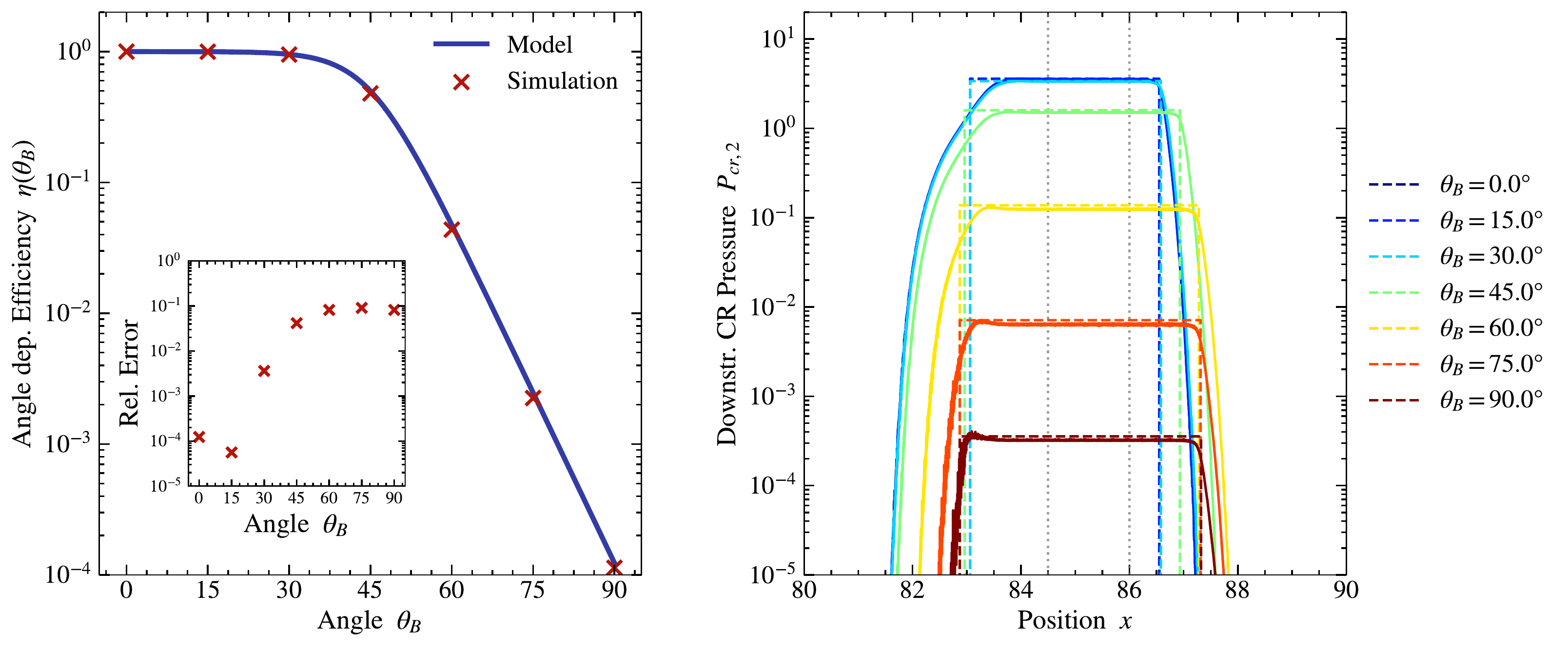}
    \caption{Tests for the accuracy of the magnetic field geometry dependent injection. \textit{Left:} Functional form of the dependency parameter and simulation data taken from the mean values of the post shock region indicated by vertical dotted lines in the RHS plot. The inset plot shows the corresponding relative errors. \textit{Right:} CR proton pressure component in the post-shock region. Colors indicate the ideal angle $\theta_\mathrm{B}$ between shock normal and magnetic field vector. Dashed lines show the analytic solution, while solid lines show the simulation output. Values of all particles are shown.}
    \label{fig:theta_B_postshock}
\end{figure*}

To test how accurately we can model the magnetic field angle dependent acceleration model we followed the approach by \citet{Dubois2019} and set up a series of shock tube tests with negligible, but constant magnetic field at a given angle to the shock propagation.
This allows us to capture the angle $\theta_\mathrm{B}$ between $\hat{n}_s$ and $\mathbf{B}$, while avoiding a kinetic impact of the magnetic field on the development of the shock.
The results of these test for the proton component can be seen in Fig.~\ref{fig:theta_B_postshock}.
The blue line in the l.h.s. of the figure shows Eq.~\ref{eq:b-angle-dependence} with $\theta_{\mathrm{crit}} = \pi/4$.
The red crosses show the results of our simulation. 
We obtained these values by taking the mean value of the post-shock region indicated by the dashed vertical lines on the r.h.s.
The small inset plot shows the corresponding relative error.
The r.h.s. shows the injected CR proton pressure component in the post-shock region.
Dashed lines indicate the analytic solution, while solid lines show the values of all SPH particles containing injected CRs.
Colors correspond to the angle between the shock normal $\hat{n_s}$ and the magnetic field $\mathbf{B}$.
In general we find excellent agreement with the analytic solutions.
The solutions stay numerically stable with very low numerical noise.
\subsubsection{Spectral slope}
\begin{figure*}
    \centering
	\includegraphics[width=0.8\fullwidth]{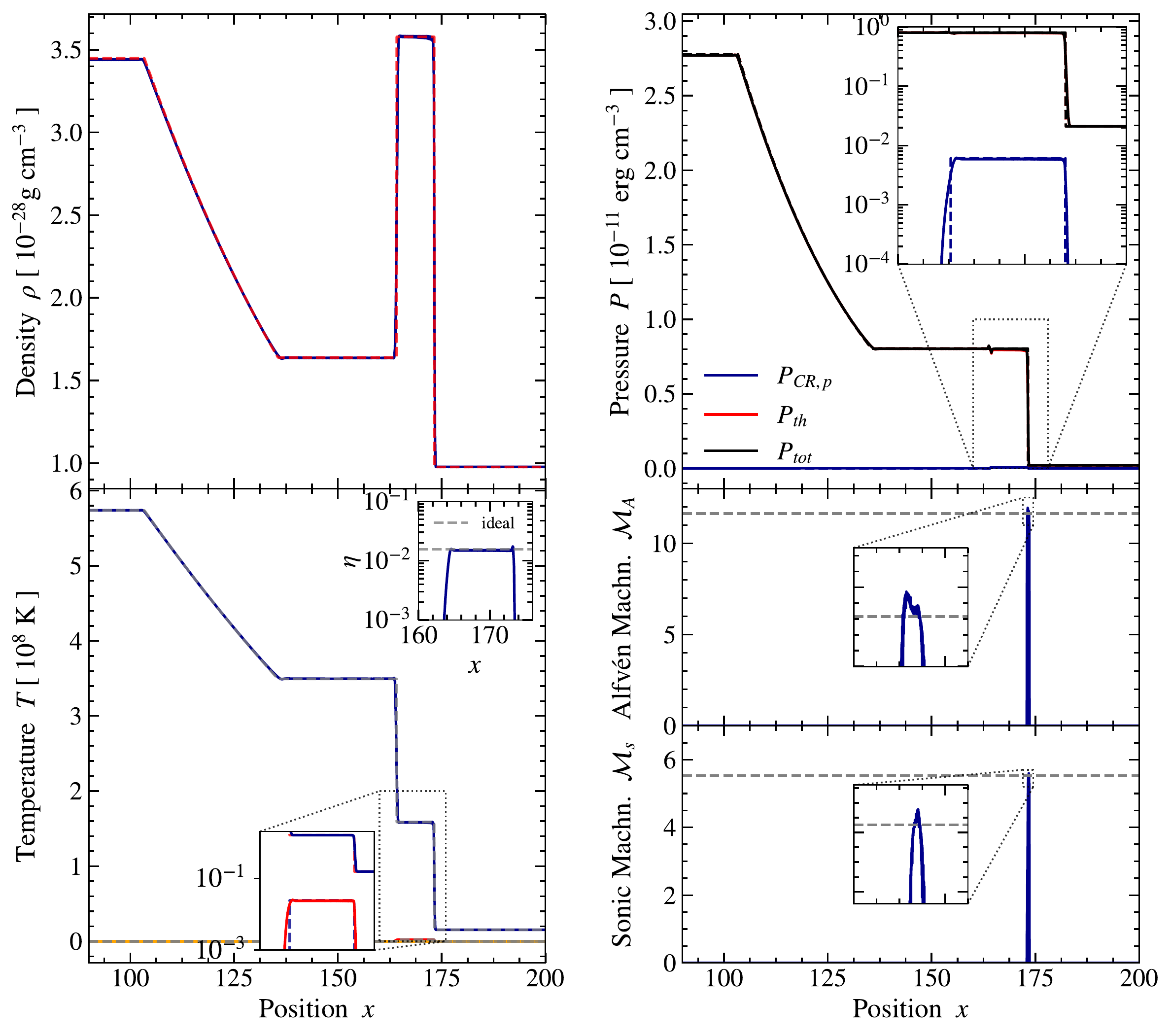}
    \caption{Final state of the shocktube test with galaxy cluster properties. Values of all particles are shown with solid lines to include numerical noise. In all plots the analytic solutions are indicated with dashed lines. \textit{Top left}: Density, \textit{Top right:} Pressure with total pressure in black lines, CR proton pressure in blue lines and thermal pressure in red lines. The inset plot shows a zoom-in on the injection region with logarithmic scaling. \textit{Bottom left:} Temperature and CR energy. The upper inset plot shows the energy ratio in the injection region, while the lower one gives a zoom-in on the injection region with logarithmic scaling. \textit{Bottom right:} Alfvén- and sonic Mach number. Inset plots zoom in on the peak of the shock. The horizontal dashed line shows the analytic solution.}
    \label{fig:shocktube_cluster}
\end{figure*}

\begin{figure}
	\centering
	\includegraphics[width=0.8\columnwidth]{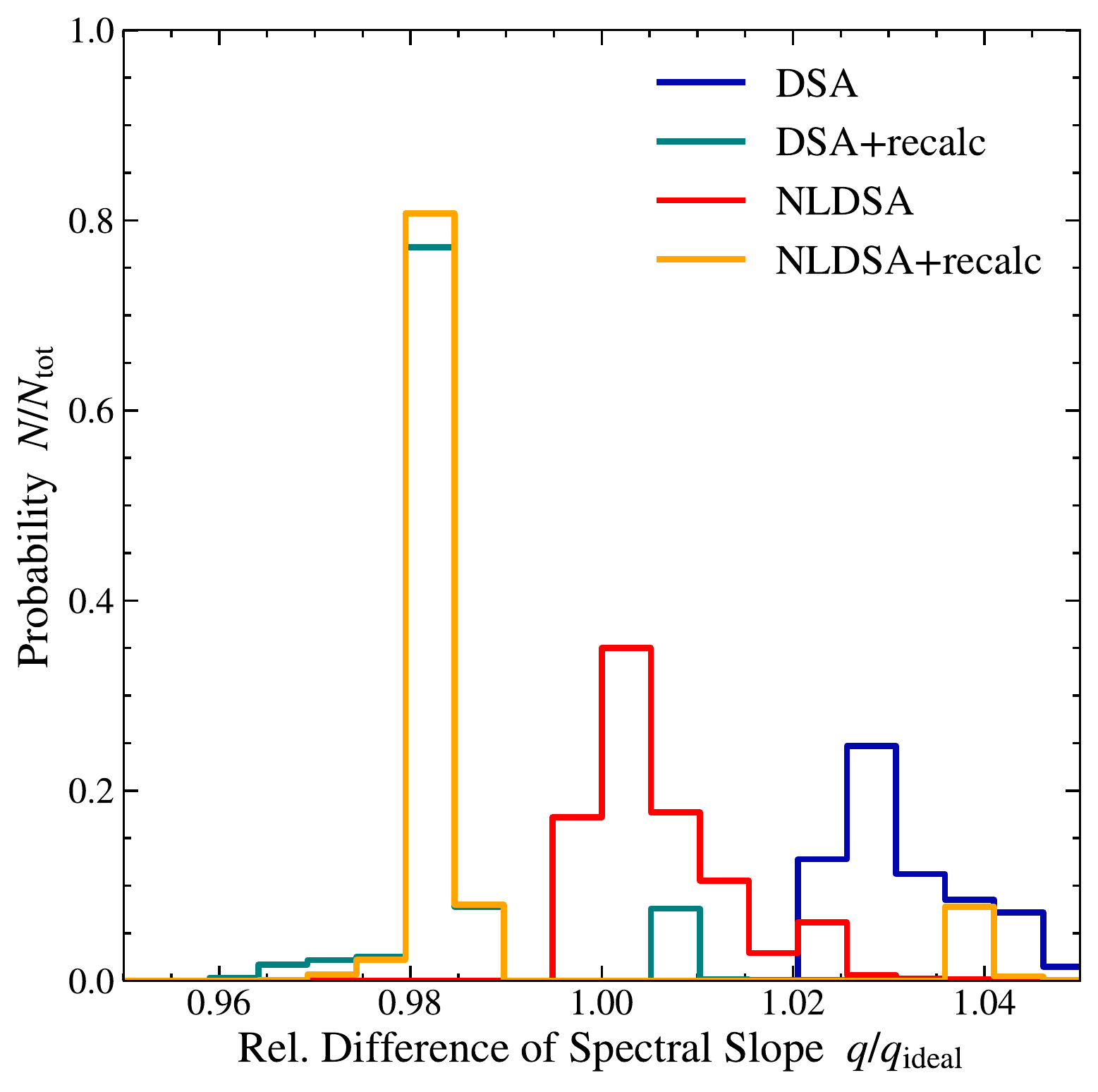}
	\caption{Histograms of the ratio between injected and ideal spectral slopes. Colors represent standard DSA and NLDSA description with and without recalculation of the shock compression ratio.}
    \label{fig:q_hist}
\end{figure}

As the shock front is smoothed out by the SPH kernel we systematically under-predict the compression ratio of the shock.
Since the velocity jump is equally smoothed out the two effects cancel out and the error of the Mach number estimate at the shock center is on a sub-percent level \citep[see][]{Beck2016b}. 
To remedy this behaviour we optionally recalculate the shock compression ratio based on the Mach number from the Rankine–Hugoniot conditions as 
\begin{equation}
    x_s = \frac{ ( \gamma + 1 ) \mathcal{M}_s^2}{(\gamma - 1 ) \mathcal{M}_s^2 + 2}
    \label{eq:shock_compress_recalc}
\end{equation}
where $\gamma = 5/3$ is the adiabatic index of an ideal gas and $M_s$ is the Mach number of the shock. This approach holds only with a small CR component and is therefore only justified for usage in structure formation shocks where the CR pressure components is expected to be small (as discussed above) and not e.g. in resolved ISM simulations with SNe, where the CR pressure component can be a significant fraction of the total pressure \citep[e.g.][]{Beck2015} and will therefore modify the shock properties.\\
For testing the accuracy of capturing the correct slope of the injected spectrum we set up a series of shock tubes with properties similar to those found in galaxy cluster shocks. Table \ref{tab:shocktubes} gives the properties of the shock initial condition and Fig.~\ref{fig:shocktube_cluster} shows the result of the simulation. As can be seen the quantities agree nicely with the analytic solution and the capture of the Alfvén Mach number and with that the capture of the Alfvén speed needed for the non-linear correction to DSA agrees very well with the analytic solution. We ran four different simulations with each DSA, DSA plus recalculation of compression ratio according to Eq.~\ref{eq:shock_compress_recalc}, NLDSA and NLDSA with recalculation. We then compared the obtained injection slopes with the ideal slopes in Fig.~\ref{fig:q_hist}. The recalculation shows promising results, as it is less broadened and in the case of DSA more accurate. For NLDSA recalculation introduces a larger error, but nonetheless stays less broadened.
We therefore accept this discrepancy for now.

\section{Cluster Merger Simulations}
\label{sec:cluster_mergers}
Idealized galaxy cluster mergers have been studied previously with great success to model X-ray emission of dynamical clusters \citep[e.g.][]{Donnert2017} and to study the origin of observed cold fronts \citep[e.g.][]{Springel2007, ZuHone2010, ZuHone2013, Walker2017} as well as velocity structures in merging clusters \citep[][]{Biffi2022}, or to model radio observations from relics \citep[e.g.][]{Weeren2010, Weeren2011, Lee_2020, Lee_2022} and secondaries \citep[e.g.][]{ZuHone2013, Donnert2014}. For a recent review on GC merger simulations see \citet{ZuHone2022}.
To test our model in a more realistic test case we ran a series of idealized galaxy cluster mergers following the best fit parameters for \textsc{CIZA J2242.4+5301-1} obtained in \citet{Donnert2017}. Specifically we use the high Mach number scenario of the Red model, which gives the Mach number closest to that obtained by radio observations, while also matching the X-ray observations. This allows us to compare the result directly to their work and well studied radio observations of the sausage relic \citep[e.g.][]{Stroe2013, Stroe2014, Stroe2016, Gennaro2018, Weeren2019}.
All these simulations are non-radiative, run with \textsc{OpenGadget3} using the improvements to SPH presented in \citet{Beck2016}, higher order $C_6$-kernels with 295 neighbors, non-ideal MHD \citep[][]{Dolag2009a, Bonafede2011}, on-the-fly shock finder \citep[][]{Beck2016b} and thermal conduction \citep[][]{Jubelgas2004, Arth2014}.

\subsection{Initial Conditions}
%
To construct the initial conditions for the galaxy cluster merger we employ a slightly modified version of the \textsc{toycluster} code \citep[see][for details]{Donnert2014a} with improvements presented in \citet{Donnert2017}. The code sets up DM and gas spheres for galaxy clusters and places them on a colliding orbit.
For the purpose of this work we will only outline the key components of the IC setup here and refer the interested reader to the aforementioned papers.
\textsc{toycluster} uses rejection sampling to set up the positions of equal-mass DM particles following a NFW profile
\begin{equation}
    \rho_\mathrm{DM} = \frac{\rho_{0,\mathrm{DM}}}{\frac{r}{r_s} \left( 1 + \frac{r}{r_\mathrm{s}} \right)^2 } \left( 1 + \frac{r^3}{r_{\mathrm{sample}}^3} \right)^{-1}
\end{equation}
where $\rho_{0,\mathrm{DM}}$ is the central DM density, $r_s$ is the NFW scale radius and $r_{\mathrm{sample}}$ is the sample radius for the DM distribution. We employ $r_{\mathrm{sample}} = 1.7 r_{200}$ \citep[see Sec 3.1 in][for a discussion about the choice of this value]{Donnert2017}.
The corresponding particle energies and from that the velocities to obtain a stable halo are then found by sampling from the particle distribution function $f(E)$. With the added complexity of an embedded gas halo within the DM halo $f(E)$ must be obtained by numerically solving the Eddington equation \citep[][]{Eddington1916}.
\begin{equation}
    f(E) = \frac{1}{\sqrt{8} \pi} \int\limits_0^E \frac{d\Psi}{\sqrt{E - \psi}} \frac{d^2 \rho}{d\Psi^2}
\end{equation}
where $\psi$ is the potential energy, $E$ the kinetic energy and $\rho$ the total density profile. 
The positions of the gas particles are found with a weighted Voronoi tessellation method \citep{Diehl2012, Arth2019}. This method defines a maximum density as a function of position, in this case a $\beta$-model \citep{Cavaliere1976} 
\begin{equation}
    \rho_\mathrm{gas}(\mathbf{x}) = \rho_{\mathbf{0},\mathrm{ICM}} \left( 1 + \frac{r^2(\mathbf{x})}{r_\mathrm{core}^2} \right)^{-\frac{3}{2} \beta}  \left( 1 + \frac{r^3(\mathbf{x})}{r_{\mathrm{cut}}^3} \right)^{-1}
\end{equation}
where $\rho_{\mathbf{0},\mathrm{ICM}}$ is the central ICM density, $r_\mathrm{core}$ is the core radius and $r_{\mathrm{cut}}$ is the cut-off radius of the gas-halo sampling.
Initially we sample a Poisson-distribution. The actual density at the particle position $\mathbf{x}$ is then found with a SPH loop and from that a displacement for the particle can be computed which will lead to a better agreement to the analytic density model. That process is repeated until the error between analytic and SPH density is below 5 per cent.
Finally, the gas temperature and from that the internal energy of particles is found from calculating the hydrostatic equilibrium temperature
\begin{equation}
    T(r) = \frac{\mu m_\mathrm{p}}{k_\mathrm{B}} \int_{r}^{R_\mathrm{max}} dr' \: \frac{\rho_\mathrm{gas}(r')}{r'^2} M_\mathrm{tot}(<r') 
\end{equation}
where $\mu \approx 0.6$ is the mean molecular mass of the ICM plasma and $k_B, m_p$ are Boltzmann constant and proton mass, respectively. To model cool-core and non-cool-core clusters $r_\mathrm{core}$ is set to $r_\mathrm{core} = r_s/9$ for cool-core and $r_\mathrm{core} = r_s/3$ for non-cool-core models \citep[see][]{Donnert2014}.\\
A final parameter is the in-fall velocity of the merging clusters as a function of the energy contained in the orbits ($X_E$) if the clusters are at rest at an infinite distance. In this parametrisation $X_E = 1$ is the maximum energy available to the system and $X_E = 0$ would mean the clusters are at rest if they are placed so that their virial radii $r_{200}$ touch. For the current work we employ the high Mach number scenario with $X_E = 0.5$ and the parameters of the Red model from \citet{Donnert2017}, which we sampled with $10^7$ gas and DM particles each. This leads to a mass resolution of $m_\mathrm{gas} = 5.1 \cdot 10^7 M_\odot$ and $m_\mathrm{DM} = 2.1 \cdot 10^8 M_\odot$ with a gravitational softening of $\epsilon = 3.4 \mathrm{kpc}$.
\begin{figure*}
	\includegraphics[width=\fullwidth]{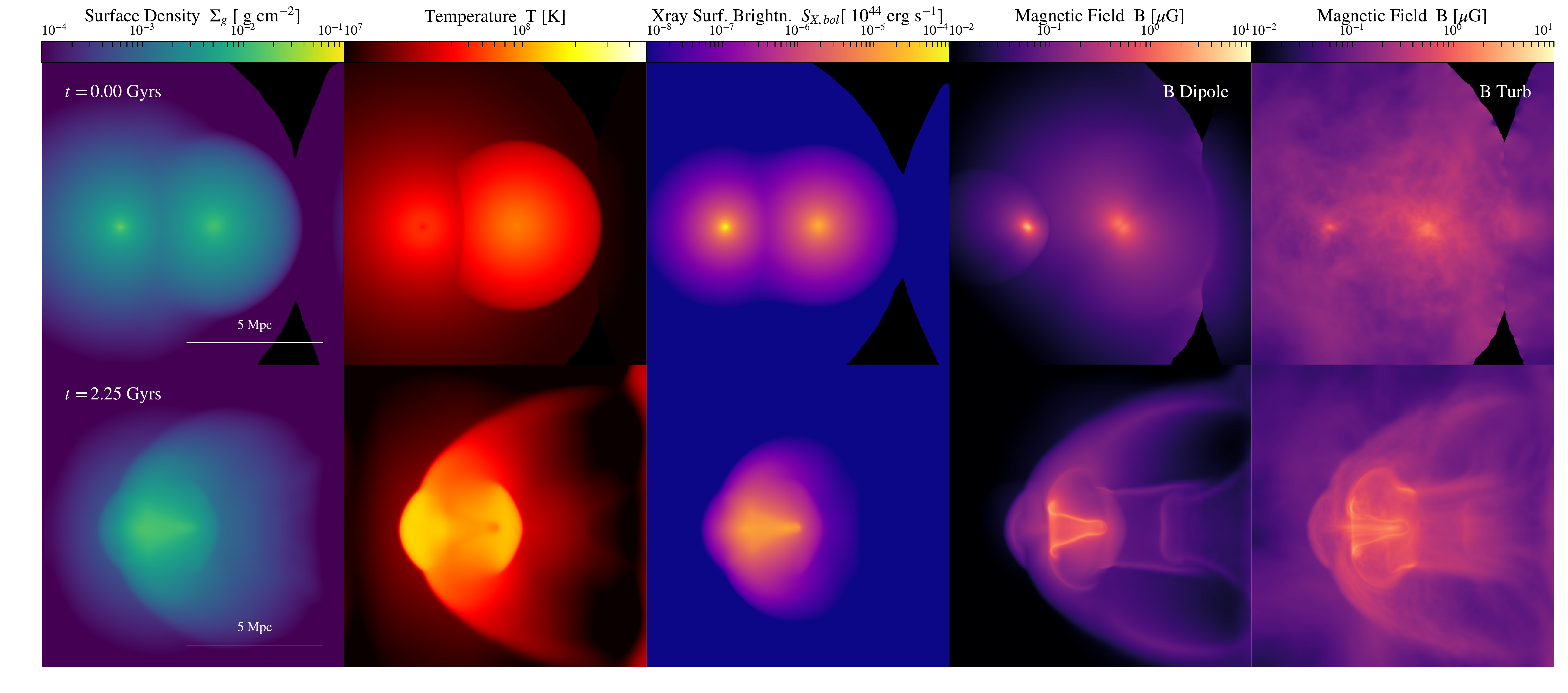}
    \caption{\textit{Upper panels:} Initial conditions for the galaxy cluster mergers. From left to right we show gas surface density, temperature, bolometric X-ray surface brightness, magnetic field strength of the dipole setup and magnetic field strength of the turbulent setup. \textit{Lower panels:} Simulation output at $t=2.25$ Gyrs as both shock waves have developed. This output time was used to obtain the results for the shock front and CR component analysis. Left and right shock correspond to northern and southern relic, respectively. The three left panels show the gas quantities of the B Dipole run. For all runs the plasma-beta is $\beta \gg 1$, which leads to a negligible impact on the gas dynamics from the magnetic field.}
    \label{fig:ic_maps}
\end{figure*}
\subsection{Magnetic Field Models}
We employ two magnetic field configurations: A dipole field and a turbulent field. For the dipole field we used the standard configuration of \textsc{toycluster} to set up a divergence free magnetic field from a vector potential. Here we follow the magnetic field model by \citet{Bonafede2011} and define a vector potential as
\begin{equation}
    \mathbf{A}(\mathbf{x}) = A_0 \left( \frac{\rho(\mathbf{x})}{\rho_0} \right)^\eta
\end{equation}
with $A_0 = 5 \mu \mathrm{G}$ as the central field strength and $\eta=0.5$ as the scaling parameter. We then compute the magnetic field components by explicitly solving the curl of the vector potential over the neighboring SPH particles.\\
For the turbulent magnetic field we set up a power spectrum in Fourier space with an amplitude $P(k) \propto k^\alpha$, where $\alpha = - 11/3$. We then sample randomly from this spectrum on a 3D grid and transform this grid into real space. In real space we can then normalize the magnetic field to the desired field strength, again $5 \mu \mathrm{G}$ and apply a density weighting as in the previous case. The normalized B-field grid is then again transformed into Fourier space for divergence cleaning, following the method described in \citet{Ruszkowski2007}.
The divergence free grid is then transformed back into real space. From there the magnetic field can be mapped to the SPH particles by Nearest Grid Point interpolation.
Both these methods result in small values for $\nabla \cdot \mathbf{B}$ with a mean relative divergence of $\vert \nabla \cdot \mathbf{B} \vert h_i / \vert \mathbf{B} \vert \approx 10^{-6}$ in the case of the dipole setup and $\vert \nabla \cdot \mathbf{B} \vert h_i / \vert \mathbf{B} \vert \approx 10^{-4}$ in the case of the turbulent setup over the course of the simulations. We find that these values are acceptable for our simulation efforts.
\subsection{Simulations}
\begin{table*}
    \caption{Configuration of the different simulation runs. From left to right we list the name of the run. Whether it was run with the dipole magnetic field setup or the turbulent magnetic field setup. Which Mach number dependent acceleration model was used. If the magnetic field geometry dependent efficiency was used. How $p_\mathrm{inj}$ was defined. How the ratio between CRp and CRe injection $K_\mathrm{ep}$ was set. And if it was run with turbulent re-acceleration.}
    \begin{tabular}{|l|c|c|c|c|c|c|c|c|}
        Model & B Dipole & B Turb & $\eta(\mathcal{M}_s)$ & $\eta(\theta_B)$ & $p_\mathrm{inj}$ & $K_\mathrm{ep}$ & $\alpha$ \\
        \hline
        KR13$d \theta_B$ & \checkmark & & \citet{Kang2013} & \checkmark & 0.1 & 0.01 & 0 &\\
        KR13$t$ &  & \checkmark & \citet{Kang2013} & & 0.1 & 0.01 & 0 &\\
        KR13$t \theta_B$ & & \checkmark & \citet{Kang2013} & \checkmark & 0.1 & 0.01 & 0 &\\
        Ryu19$t$ & & \checkmark & \citet{Ryu2019} &  & 0.1 & 0.01 & 0 &\\
        Ryu19$t \theta_B$ & & \checkmark & \citet{Ryu2019} & \checkmark & 0.1 & 0.01 & 0 &\\
        Ryu19$t \theta_B p_\mathrm{inj}$ & & \checkmark & \citet{Ryu2019} & \checkmark & Eq.~\ref{eq:p_inj} & Eq.~\ref{eq:Kep} & 0 &\\
        Ryu19$t \theta_B p_\mathrm{inj} q_\alpha$ & & \checkmark & \citet{Ryu2019} & \checkmark & Eq.~\ref{eq:p_inj} & Eq.~\ref{eq:Kep} & Eq.~\ref{eq:new_dsa_slope} &\\
    \end{tabular}
    \label{tab:models}
\end{table*}
We ran a total of seven different simulations to study the impact of the different components of our model. We summarize the runs in Tab.~\ref{tab:models} and will give a brief overview over the different setups and the naming convention, as well as their motivation in this section.
First we distinguish between the different Mach number dependent acceleration efficiency models. For these runs we use the models by \citet{Kang2013} and \citet{Ryu2019}, denoted by KR13 and Ryu19 respectively.
The most simple run is KR13$t$ with only the sonic Mach number dependent acceleration efficiency employed, a fixed injection momentum $\hat{p}_\mathrm{inj} = 0.1$ and a fixed electron to proton injection of $K_\mathrm{ep} = 0.01$ \citep[as in e.g.][]{Hong_2015}. We use this as a baseline to see the impact of a Mach number dependent efficiency model and use the magnetic field only for the synchrotron analysis in Sec.~\ref{subsubsec:synch_morphology}.
Next we keep the previous parameters and include shock obliquity dependent acceleration efficiencies. We test this for the ordered, dipole magnetic field and the turbulent magnetic field in KR13$d \theta_B$ and KR13$t \theta_B$ respectively. A visualisation of the intial conditions with dipole and turbulent magnetic field can be seen in the two upper right panels of Fig.~\ref{fig:ic_maps}.
We then switched to the \citet{Ryu2019} efficiency model where we use the turbulent setup to first test only the effect of switching to this more modern Mach number dependent efficiency model in Ryu19$t$ and then include magnetic field geometry dependent acceleration in Ryu19$t \theta_B$.
The next simulation again uses the more modern Ryu19 efficiency, shock obliquity dependent injection and on-the-fly calculation of $\hat{p}_\mathrm{inj}$ and $K_\mathrm{ep}$. We use this to study how our model behaves with a more modern injection efficiency and more complex parameter combinations for the distribution functions in run Ryu19$t \theta_B p_\mathrm{inj}$.
Last, we reuse all settings from the previous simulation, but also include the computation of a slope based on non-linear DSA following Eq. \ref{eq:new_dsa_slope}.
All simulations were run with CR distributions in the range $\hat{p} \in [10^{-1}, 10^5]$. This represents the full range of the spectrum in the case of a fixed $\hat{p}_\mathrm{inj}$ and makes it easy to compare these results to the simulations with an on-the-fly calculation of $\hat{p}_\mathrm{inj}$.
However this puts strain on our approximation $T(p) \approx pc$, as particles with $\hat{p} \approx 10^{-1}$ can not be considered ultra-relativistic and the transition between $\gamma = \frac{4}{3}$ and $\gamma = \frac{5}{3}$ occurs around $\hat{p} \approx 10^{-1} - 10^1$ \citep[see Fig. 2 in][]{Girichidis2022}.
This leads to our pressure estimates being a lower limit.\\
We resolve the CR proton spectrum with 12 bins (2 bins/dex) and the electron spectrum with 96 bins (16 bins/dex) for each of our $10^7$ resolution elements. 

\subsection{Shock Fronts \label{sec:shock_front}}
\begin{figure*}
	\centering
	\includegraphics[width=0.9\fullwidth]{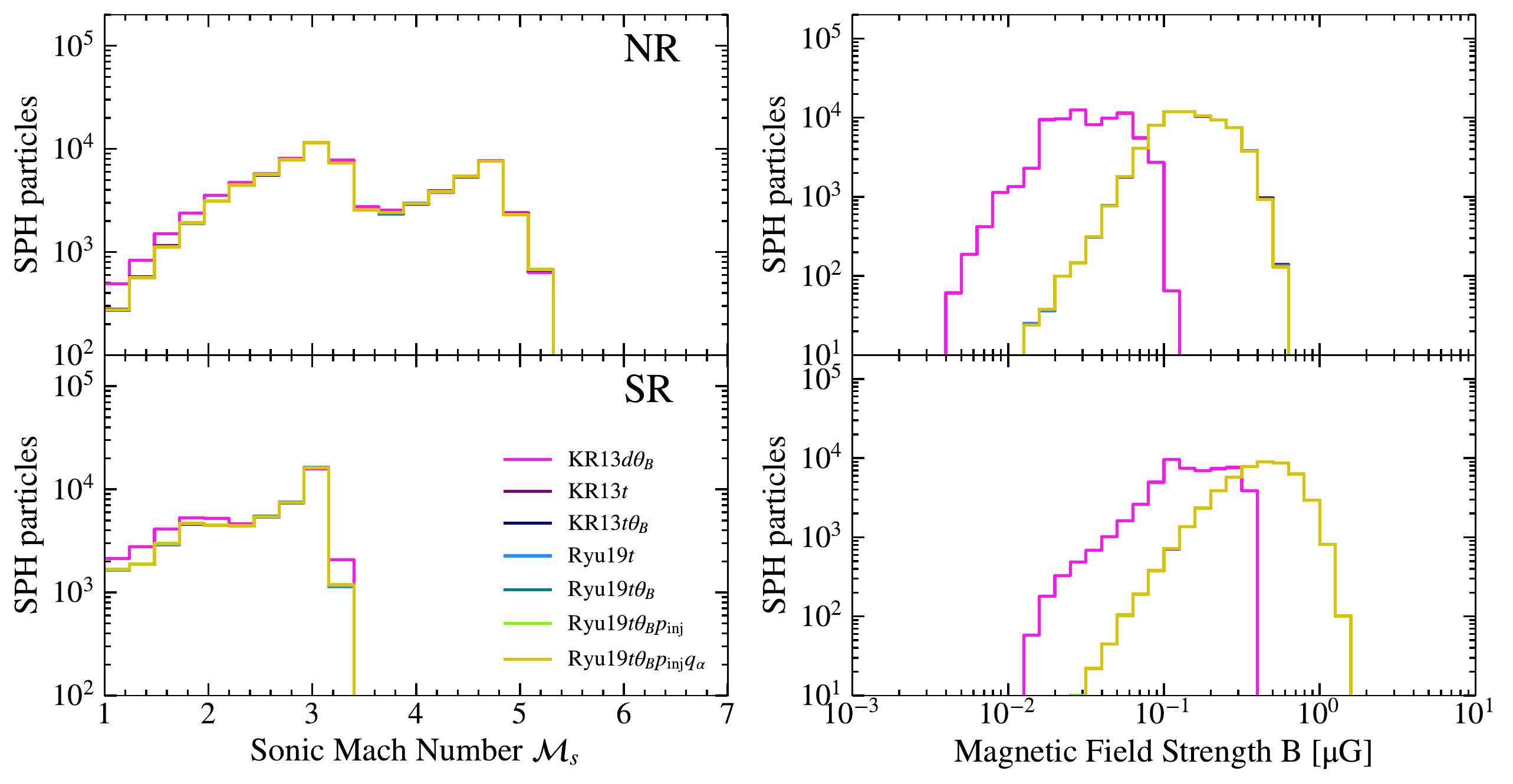}
	\caption{Shock properties of the northern relic (NR, top panels) and southern relic (SR, lower panels) in our simulations. \textit{Left:} Histograms of sonic Mach number distribution of all shocked SPH particles. \textit{Right:} Magnetic field strength distribution of all shocked particles. . The magnetic field strength of the NR is significantly below common values of $2 - 5 \mu G$ commonly found in cluster shocks \citep[see tables in][]{Weeren2019}. As the SR expands into the wake of the larger cluster the magnetic field is stronger.}
    \label{fig:shock_statistic}
\end{figure*}
For the rest of the paper we will study the shocks moving along positive and negative $x$-direction. Here the left moving shock corresponds to the northern relic (NR) and the right moving shock to the southern relic (SR) in \citet{Donnert2017}. 
We show the state of the simulation at $2.25$ Gyrs, which we will use for the following analysis of the NR in the bottom panels of Fig.~\ref{fig:ic_maps}.
For the analysis of the SR we use an earlier snapshot at 1.96 Gyrs, as at the later time the shock has already extended further into the track of the larger cluster and has been deformed by boundaries of this track.
Fig.~\ref{fig:shock_statistic} gives a histogram of sonic Mach number and magnetic field strength distribution in the shock fronts.
As expected the Mach number distribution does not differ significantly between the runs. Even for the run with the highest efficiency (KR13$t$) does not inject enough CRs to significantly alter the downstream equation of state and with that the shock speed. This makes for good comparisons of the different simulations.
A larger discrepancy can be found in the distribution of magnetic field strengths of the shocked particles, where the KR13$d \theta_B$ model shows half dex lower magnetic field strengths. As the shock is detected slightly ahead of the density and temperature jump the magnetic field amplification by the shock is not completed in the shocked particles either. This means that the shocked particle primarily probe the upstream magnetic field, which is stronger in the turbulent setup.
In addition to that we find that the downstream magnetic field is lower in our simulation, compared to typical observational values of $2 - 5 \mu G$ commonly found in cluster shocks \citep[see tables in][]{Weeren2019}.
We will discuss the implications of this for the synchrotron emission we obtain directly from the electron population in our particles in Sec.~\ref{sec:synch_power}.

\subsection{Synchrotron Emission}

One of the key advantages of a spectral CR model is the possibility to obtain the synchrotron emission of the population directly from the simulation output.
To calculate the synchrotron emission of our distribution function we take the same approach as \citet{Donnert2016, Mimica_2009} and follow \citet{Ginzburg1965}. With this the synchrotron emissivity $j_\nu$ in units of erg cm$^{-3}$ s$^{-1}$ Hz$^{-1}$ for an distribution function of CR electrons in dimensionless momentum space $f(\hat{p}, t)$ can be expressed as 
\begin{align}
    j_\nu(t) 
    &= \frac{\sqrt{3} e^3}{c} \: B(t) \: \sum\limits_{i=0}^{N_\mathrm{bins}} \:\int\limits_0^{\pi/2} d\theta  \sin^2\theta \:  \int\limits_{\hat{p}_\mathrm{i}}^{\hat{p}_\mathrm{i+1}} d\hat{p} \:\: 4\pi \hat{p}^2 f(\hat{p}, t) \: K(x)
    \label{eq:synch_emissivity}
\end{align}
where  $e$ is the elementary charge of an electron, $c$ is the speed of light, $\hat{p}$ is the dimensionless momentum and $K(x)$ is the first synchrotron function
\begin{equation}
    K(x) = x \int_x^{\infty} dz \ K_{5/3}(z)
    \label{eq:synch_kernel}
\end{equation}
using the Bessel function $K_{5/3}$ at a ratio between observation frequency $\nu$ and critical frequency $\nu_c$
\begin{equation}
    x \equiv \frac{\nu}{\nu_c} = \frac{\nu}{C_\mathrm{crit}  B(t) \sin\theta \: \hat{p}^2}; \quad C_\mathrm{crit} = \frac{3e}{4\pi m_e c} \quad .
\end{equation}
We solve the momentum integrals by employing the Simpson rule, which constructs a mid-point by interpolating the simulated spectrum and the pitch angle integrals with a trapez integration.

\section{The Northern Relic}
\label{sec:NR}
\begin{figure*}
    \centering
    \includegraphics[width=0.95\fullwidth]{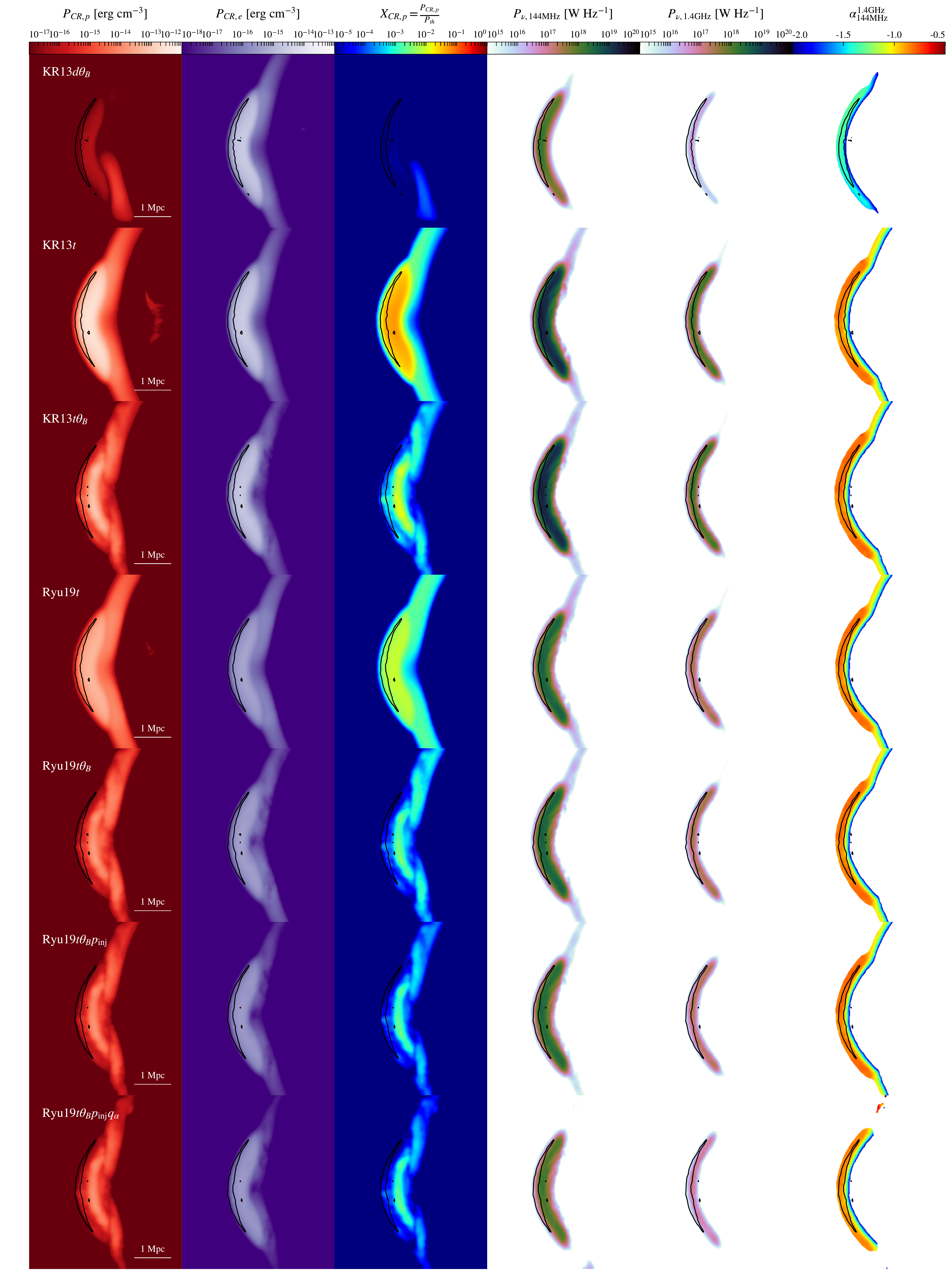}
    \caption{Simulation results for the northern relic (NR). We show, from left to right, the proton and electron pressure components, the ratio between proton and thermal pressure, synchrotron power at 144 MHz, synchrotron power at 1.4 GHz and the spectral slope obtained by a single powerlaw fit between those frequencies. The images contain the average values in a 200kpc slice, centered on the shock. The black contours show the SPH particles in the shock with a sonic Mach number $\mathcal{M}_s > 4.5$ which corresponds to the center of the shock.}
    \label{fig:cr_maps_NR}
\end{figure*}
First we will focus on the CR pressure component, synchrotron emission and the time evolution of the spectral distributions of protons and electrons of the northern relic (NR). We will discuss the southern relic in the next section.

\subsection{Injection}

We show the injected pressure component of CR protons and electrons, as well as the ratio between proton and thermal pressure in the first three columns of Fig.~\ref{fig:cr_maps_NR}.
As discussed above, our approximation of $T(p) \approx pc$ only provides a lower limit here.\footnote{We also note that for the electrons Coulomb losses at the lowest end of our distribution function are efficient enough to cool away a substantial amount of the total energy density and it would be more consistent to only show the energy density of the synchrotron and IC dominated part of the spectrum. However we show the total pressure component here to illustrate the difference in injection, which would be less visible if only the fast-cooling part of the spectrum was considered.}
The top panels show the results for the model KR13$d \theta_B$. In the dipole magnetic field case the shock expands into a lobe of the dipole setup, causing a very oblique magnetic field geometry over the entire shock surface. This strongly suppresses the injection of CR protons. In the electron case this leads to a preferential acceleration and with that a smoothly injected electron component whose energy density surpasses that of the protons. As a result of this the ratio between CR proton and thermal pressure is negligible and proses no stress on the observational constraints.
In the KR13$t$ model, which does not use the shock obliquity dependent acceleration efficiency, we find a smooth injection of both components and a energy ratio which strictly follows the fixed $K_\mathrm{ep} = 0.01$.
The CR proton to thermal pressure ratio is significantly higher than in the previous run at 5\% and with that surpasses the observational limit, which we expect to become a problem with a more realistic setup that also includes multiple shocks and the more efficient re-acceleration models.
Including $\eta(\theta_B)$ in the run KR13$t \theta_B$ results in a more varying CR component behind the shock, for protons. 
We can see a decreases in proton pressure where the shock propagated through regions of varying shock obliquity and find a structure behind the shock with increased proton pressure.
This is also evident in the map of thermal to CR proton pressure ($X_{\mathrm{CR},p}$), where the maximum ratio behind the shock, caused by a pocket of perpendicular magnetic field does approach the observational limits, but there are also regions of significantly lower CR proton pressure.\\
Once we switch to the most modern Ryu19 efficiency models we always obtain CR proton pressure components in agreement with the observational limits.
The Ryu19$t$ run shows smooth injection at the shock front for both electrons and protons, as expected with a injection scheme only dependent on Mach number.
The values of $X_{\mathrm{CR},p}$ stay below 1\%, however in a full cosmological simulation with multiple shocks and re-acceleration this picture could change.
Once the magnetic field angle dependent acceleration efficiency in Ryu19$t\theta_B$ is switched on however, CR proton acceleration is suppressed further, making this our favored model for CR proton acceleration.
Including on-the-fly calculation of $\hat{p}_\mathrm{inj}$ and NLDSA slopes does not significantly alter this picture. The CR proton to thermal pressure ratio lies well below observational limits. This is for one caused by the lower efficiency and for another by the fact that not all of the energy available for injection is represented by our CR population and therefore remains in the thermal gas component.
Nevertheless we expect this efficiency model to behave like in other studies in a full cosmological simulation and suppress acceleration and re-acceleration of CR protons enough to be consistent with observations.

\subsection{Radio Relic Morphology}
\label{subsubsec:synch_morphology}
We applied the calculation of synchrotron emissivity to the CRe populations injected at the bow shock. After calculating the emissivity per particle we mapped the particles to a 2D image, following the algorithm described in \citet{Dolag2005}. We do not smooth the images with a radio beam to retain the intrinsic information, for simplicity.
The result can be seen Fig.~\ref{fig:cr_maps_NR}. For this section we will address the morphology of the synchrotron emission shown in columns four and five.\\
In all cases the 144 MHz emission shown in the fourth panels still closely follows the total CRe pressure component, albeit we can see that the absolute emission behind the shock is decreasing by more than 2 orders of magnitude due to the cooling of the electron population.
This is especially evident in the KR13$d \theta_B$ run where the magnetic field is significantly smoother behind the shock, indicating that the decrease in emission is mainly caused by the cooling electrons. 
With the turbulent magnetic field models we see the imprints of the turbulent field behind she shock in the low frequency emission for models KR13$t$ - Ryu19$t \theta_B p_\mathrm{inj} q_\alpha$.
As seen in the suppression of the CR proton component the shock has a predominantly large obliquity and with that favors CR electron acceleration. Since the morphology of the magnetic field in the medium the northern shock travels thorugh is not very complex however, it has little impact on the relic morphology in this case.
For the 1.4 GHz emission images in the fifth column of Fig.~\ref{fig:cr_maps_NR} we see a significantly narrower emission zone behind the shock, caused by the much shorter cooling times of the radio bright electrons at this frequency.
The run KR13$d\theta_B$ shows again very smooth synchrotron emission at the shock front, which gradually decreases and also shows a smooth structure behind the shock.
Including the turbulent setup, but switching off $\eta(\theta_B)$ gives a somewhat similar smooth synchrotron surface at the shock, as the shock travels through a fairly homogenious medium apart from the magnetic field structure.
This repeats for the remaining models.
Again including the inculsion of $p_\mathrm{inj}$ computation does not significantly alter this image.
The inclusion of the $q_\alpha$-model does decrease the synchrotron emission significantly however, as the schock is only very weakly magnetized.
This causes $\alpha$ in Eq. \ref{eq:new_dsa_slope} to approach zero and with that the slope follows the standard DSA prediction.

\subsection{Spectral Steepening}

One distinct feature of radio relics is the steepening of the synchrotron spectrum behind the estimated shock front.
This is in the literature commonly attributed to the cooling of high energy, synchrotron bright electrons due to their synchrotron emission and inverse Compton scattering off background photons \citep[][]{Weeren2019}.
This steepening has been studied with toy models \citep[e.g.][]{Donnert2017} as well as with idealized simulations \citep[e.g.][]{Stroe2016}.
In our simulations we can obtain the spectral steepening directly from the aging electron population within every resolution element.
For this we construct images by calculating the emissivity per particle and integrating along the line of sight, as described in the previous section. Taking the intensity of the same pixel at two frequencies $\nu \in [144, 1400]$ MHz and fitting a single power-law between the results gives the spectral slope of the synchrotron spectrum.
The results of this can be seen in the right panels of Fig.~\ref{fig:cr_maps_NR}.
We chose color range and map to closely resemble Fig. 4 in \citet{DiGennaro2018}.
Generally we find reasonable agreement with the spectal morphology apart from the KR13$d \theta_B$ model.
At the shock front we see a constant spectral slope of $\alpha \sim [-0.8, -1.0]$ in agreement with observations.
We note that this region is broader than the observed counterpart as this is still contained in the numerical acceleration region. For this reason a constant power-law is injected resulting in a constant spectral slope.
We note therefore that the actual spectral image should be considered starting from the center of the Mach number contour and to the right from there. 
Behind the shock we observe a clear gradual steepening up to and in principle beyond $\alpha = -2.0$. 
We chose to cut the image off at that slope for reasons of comparability to observations.
We note however that the regions of steep radio spectra $\alpha \approx -1.5 \sim -2.0$ are smaller than in observations.
Additional tests show that we can extend these zones my choosing deeper slices through the relic.
Since the shocks in these simulations are nearly perfectly bowl shaped and have a very even Mach number distribution this leads to more projection effects introduced by this only somewhat realistic setup.\\
We find that the spectral discrepancy in the KR13$d \theta_B$ run is most likely caused by the magnetic field morphology and will discuss this in more detail in the next section.
The underlying morphological resemblense to the sausage relic for all other models however gives us confidence to further study radio relic morphologies in large scale simulations of galaxy clusters with more realistic merger shocks.

\subsection{Synchrotron Surface Brightness}
\label{sec:synch_power}
\begin{figure}
    \centering
	\includegraphics[width=\columnwidth]{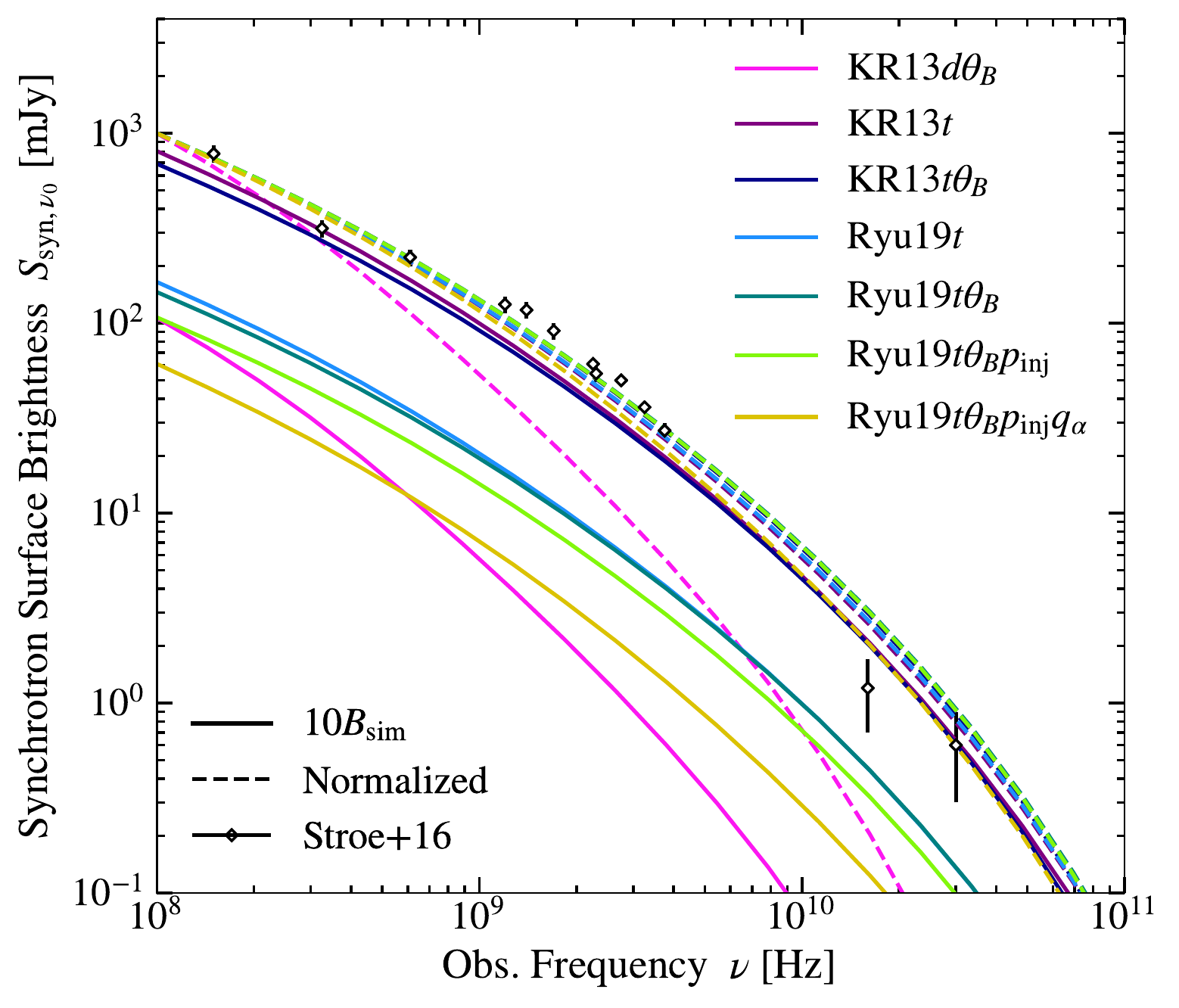}
    \caption{Synchrotron spectra as a function of observational frequency for the different models. Solid lines indicate the synchrotron power computed from 10 times the intrinsic magnetic field, to compensate for the too small magnetic field in the post-shock region. Dashed lines show the same spectra, but normalized to $1$ Jy at 100 MHz to compare spectral shapes. Diamnonds with error bars indicate the observations from \citet{Stroe2016}.}
    \label{fig:synch_spectra}
\end{figure}
We show the total synchrotron surface brightness of the NR as a function of observational frequency in Fig.~\ref{fig:synch_spectra}.
As our idealized simulation is not cosmological we assume a Planck 2018 cosmology \citep[][]{PlanckCollaboration2020} and place the relic at $z=0.1921$ for the conversion between our intrinsic radio power to observable surface brightness.
As mentioned above we find the downstream magnetic field associated with this emission is roughly one order of magnitude below observations. The resulting total synchrotron brightness is very sensitive to the magnetic field as well as the free parameters of relic volume and injection efficiency due to the shock Mach number.
To remedy this we multiply the intrinsic magnetic field with a factor of 10 and recalculate the synchrotron spectra. This is shown with the solid lines. To account for relic volume and injection discrepancy we normalized the spectra to 1 Jy at 100 MHz, the result is shown in the dotted lines. This allows us to compare the total spectral shape.
We find good agreement for the KR13$t$ and KR13$t\theta_B$ models. Both the absolute surface brightness and the shape of the spectrum match observations well. Only above 10 GHz the spectrum proves to be slightly too shallow, which can easily be attributed to the lack of synchrotron cooling due to a lower magnetic field in the simulation.
All Ryu19 models lie significantly below the observed spectrum, which follows trivially from the lower injection efficiency.
However once the spectra are normalized to 1 Jy their spectral shapes agree very well, with the observations with the Ryu19$t \theta_B p_\mathrm{inj} q_\alpha$ run showing the best agreement due to its steeper spectrum.
To rule out a systematic error in the injection we performed a shock tube test with the observational properties obtained by \citet{Weeren2010, Ogrean2014, Akamatsu2015} as used in the analytic approach by \citet{Donnert2017} to analyse the origin of discrepancy of our results. We employ the same parameters for the CR model as in the analytic work with the KR13 acceleration model, $K_\mathrm{ep} = 0.01$ and $\hat{p}_\mathrm{inj} = 0.1$. The result of this test are shown in Fig.~\ref{fig:ciza_shocktube}.
We obtain the analytic solution for density, pressure and the target Mach number with high accuracy. 
We then calculated the synchrotron emissivity per particle for a fixed magnetic field of $B = 5 \mu G$.
The results match the emissivity shown in Fig. 3 of \citet{Donnert2016} (indicated by the horizontal gray line) quite well.
This result, combined with the normalization approach above leads us to believe, that the discrepancy in the radio emission is mainly driven by the magnetic field strength.

\subsection{Spectral Evolution of a Tracer Particle}
\begin{figure*}
    \centering
    \includegraphics[width=\fullwidth]{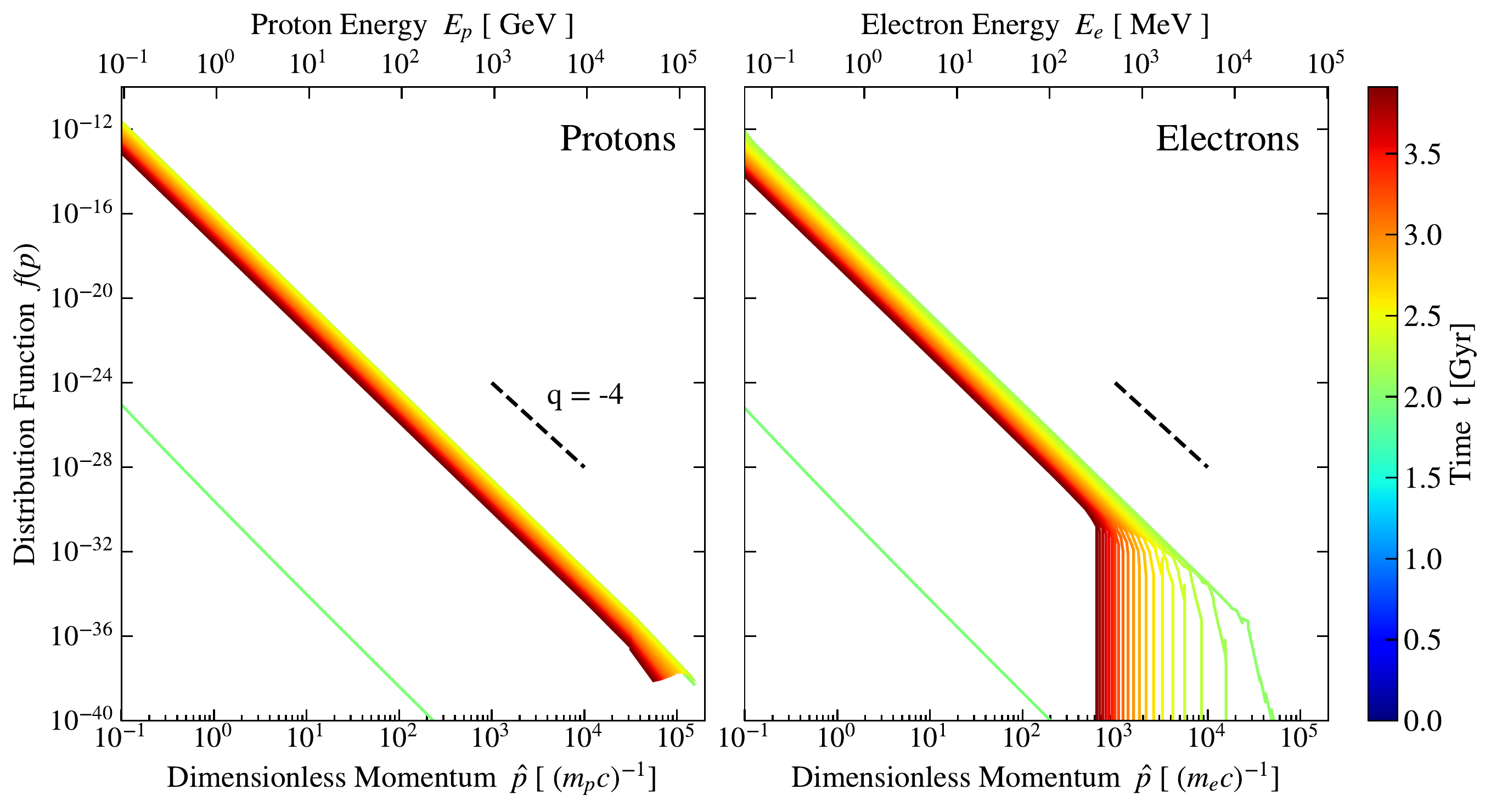}
    \caption{We show the time evolution of the proton- and electron distribution function of an arbitrarily chosen tracer particle. \textit{Left:} proton distribution function. \textit{Right:} electron distribution function. We can see the onset of the acceleration in light green and how it differs from the fully injected spectrum. The proton population shows adiabatic expansion after the injection with the spectral cutoff moving into the upper-most bin. The electron population shows adiabatic expansion and free cooling due to synchrotron emission and inverse Compton scattering. Both distribution functions stay numerically stable.}
    \label{fig:cluster_merger_spectrum}
\end{figure*}
To show the spectral evolution of our tracer particles we randomly picked one particle in the NR at $t = 1.96$ Gyrs in the Ryu19$t \theta_B p_\mathrm{inj} q_\alpha$ model, as it best reproduces the synchrotron spectrum of the sausage relic. We show the evolution of said particle in Fig. \ref{fig:cluster_merger_spectrum}.
The light green line shows the onset of the CR injection.
With the steeper slope than the later spectrum and the lower normalisation of the distribution function this suggests that the particle sat at the increasing flank of the numerically broadened shock at the output time of the snapshot.
The subsequent snapshot already sees the fully injected spectrum.
This supports the previous point of an limiting despcription in the case of a purely post-processed CR model.
A spectrum which is injected only based on the initial snapshot will significantly under-predict CR related observables for this tracer particle. 
After the injection both protons and electrons experience adiabatic expansion as they pass the shock and the gas expands again in the outskirts of the cluster. 
The electron population experiences free cooling due to synchrotron and IC losses.
The steepening of the particle distribution function is less pronounced compared to the right panel of Fig.~\ref{fig:ic_spectra} due to the flatter spectrum, attributed to the strong shock.
This shock leads to injection slopes of $q \sim -4$ and with that the fringe case of Eq.~\ref{eq:q_kardashev} where a spectrum with this slope contains its power-law shape up until the spectral cutoff.
Even in this numerically challenging case and in a production simulation we find that the spectrum stays stable. 
We note that, as discussed above, the current implementation lacks low-momentum cooling for protons and electrons and therefore over-estimates the CR pressure component at late times.
We accept this limitation for the current work, as we are interested in injection and high-momentum cooling of electrons at this point.

\section{The Southern Relic}
\label{sec:SR}
\begin{figure*}
    \centering
    \includegraphics[width=\fullwidth]{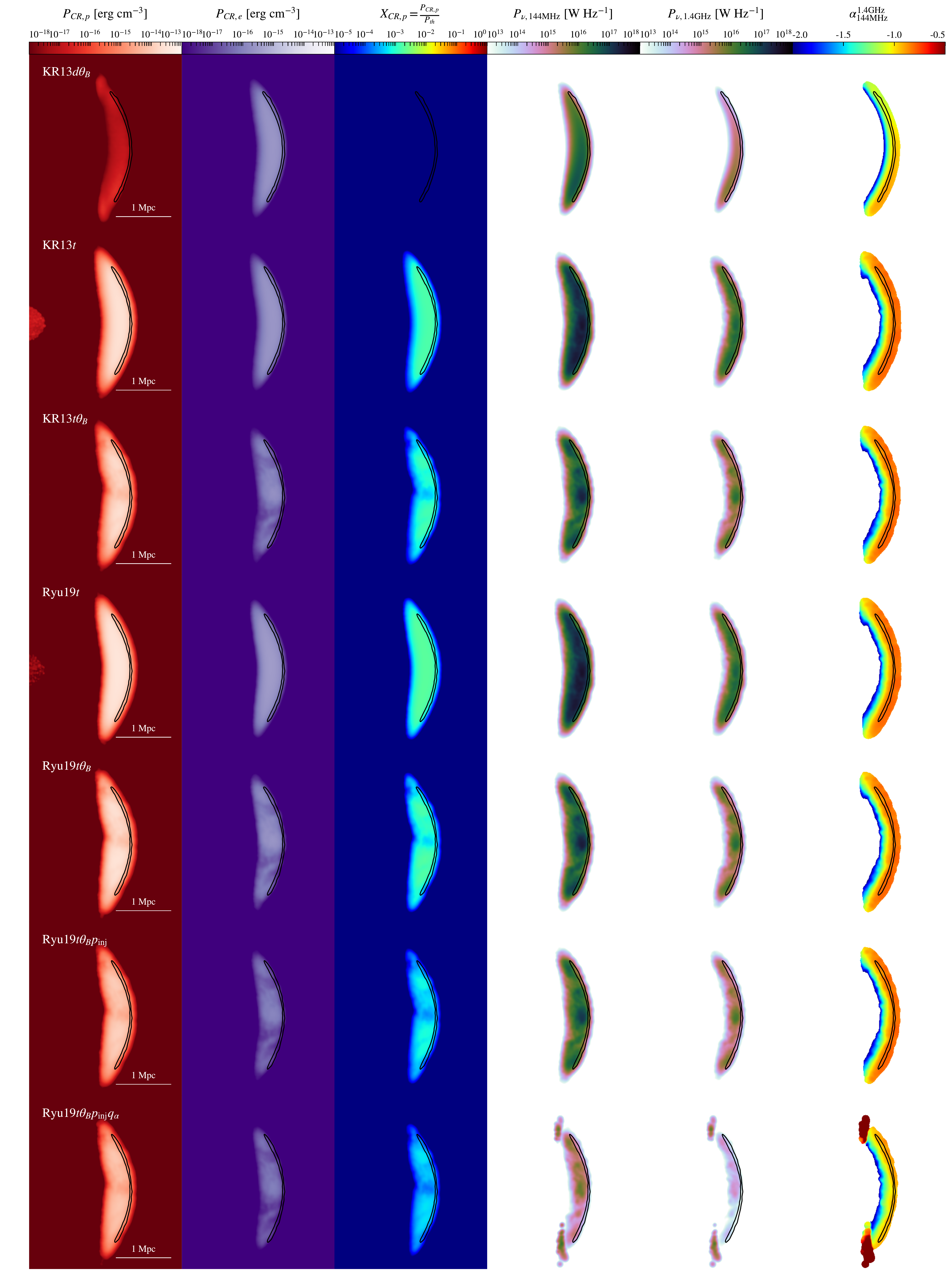}
    \caption{We show the same quantities as in Fig. \ref{fig:cr_maps_NR} for the southern relic. The black contours show the SPH particles in the shock with a sonic Mach number $\mathcal{M}_s > 3$ which corresponds to the center of the shock.}
    \label{fig:cr_maps_SR}
\end{figure*}
The southern relic (SR), the shock moving towards the right in our simulations, provides an excellent testbed for the different components of our acceleration model. Since this shock moves through a more perturbed medium, originating from the movement of the southern progenitor through the gas halo of the northern one, we can more clearly observe the impact of the different parts of the model.

\subsection{Injection}
We will again study the injected CR components first, shown in the first through third column in Fig. \ref{fig:cr_maps_SR}.\\
As was the case for the NR the shock of the SR in the KR13$d \theta_B$ run expands into a pocket of a dipole from the dipole setup leading to a strongly oblique shock.
This significantly suppresses the CR proton injection, while favoring the electron injection, which leads to a negligible CR proton to thermal pressure ratio.
In KR13$t$, the turbulent setup without $\theta_B$ computation, the injection is very smooth and again closely follows the fixed $K_\mathrm{ep} = 0.01$ ratio.
Including the obliquity dependent acceleration leads to a more complex morphology of the injected component as the efficiency fluctuates due to pockets of large and small obliquity.
We can also see how this leads to complementary behaviour between proton and electron injection in the KR13$t\theta_B$ and Ryu19$t \theta_B$ runs.
In both KR13 runs the ratio between CR proton and thermal pressure is below 1$\%$, however again this picture could change with multiple shocks and re-acceleration.
This persists in the Ryu19$t$ run, as both efficiency models lie within a factor of $\sim 2$ of each other at $\mathcal{M}_s \approx 3$.
Including $\eta(\theta_B)$ in the injection model again reduces the proton injection enough to account for this potential problem.
As was the case with the NR including $p_\mathrm{inj}$ and $q_\alpha$ computation does not change the result significantly for protons.
We find that discrepancy in CR electron injection stems mainly from the on-the-fly calculation of $K_\mathrm{ep}$, which leads to a significant decrease in this weaker shock.

\subsection{Radio Relic Morphology}
Considering the synchrotron surface brightness of the SR in the fourth and fifth panels of Fig. \ref{fig:cr_maps_SR} gives a strong insight into the effect of the different components of our acceleration model.
The KR13$d \theta_B$ run again shows a very smooth emission due to the smooth injection and the less turbulent magnetic field behind the shock.
While the injection of the CR electron component in the KR13$t$ run is similarly smooth we can see the imprint of the more turbulent magnetic field in the synchrotron emission at 144 MHz.
Including $\eta(\theta_B)$ in this setup gives rise to the complex morphology with radio brighter nodes in the radio relic which can be observed in a number of of observed examples  \citep[see][for a deeper analysis of the origin of substructure in relics]{Dominguez-Fernandez2021, Dominguez-Fernandez2021a}.
As discussed above the inclusion of $p_\mathrm{inj}$ and $K_\mathrm{ep}$ calculation in the run Ryu19$t \theta_B p_\mathrm{inj}$ reduces the CR electron injection significantly and with that reduces the synchtron emission.
The synchrotron emission is further reduced by including $q_\alpha$ and with that steepening the spectrum of the electron population, leaving fever high-energy electron to produce synchrotron emission.
In all runs the synchrotron emission is at least two orders of magnitude below that of the NR.
This puts strong observational constraints on the synchrotron brightness of these relics, providing an explanation for the occurence of single-relics, as discussed in \citet{Weeren2010}.

\subsection{Spectral Steepening}
As was the case for the NR, the SR shows significant spectral steepening following the shock propagation.
Again the Mach number contours are shown in black, indicating the center of the shock, behind which electron ageing can set on.
The KR13$t$ and Ryu19$t$ runs, with the exception of Ryu19$t \theta_B p_\mathrm{inj} q_\alpha$, show very consistent results.
In the KR13$d \theta_B$ run we find steeper spectra which we find to stem from the magnetic field morpohology, as was the case in the NR.
The steeper spectrum in the Ryu19$t \theta_B p_\mathrm{inj} q_\alpha$ run follows trivially from the steeper injection spectrum caused by our accounting for NLDSA.
Comparing the spectral morpohology with observations is difficult in this case, as the SR in \texttt{CIZA} is very dim.
This shock seems to deform the outward moving shock and could be a site of electron re-acceleration.
We will study this concept in future work with cosmological simulations, which contain more complex systems that allow for colliding shocks and multiple shock scenarios.

\section{Conclusions}
\label{sec:conclusion}

We introduced a novel implementation of an on-the-fly Fokker-Planck solver for spectrally resolved CR physics included in the cosmological Tree-SPMHD code \textsc{OpenGadget3}.
We showed that the solver performs well for test cases and reproduces CR injection at shocks, spectral changes due to adiabatic changes of the surrounding gas and high-energy cooling of CR electrons with high accuracy and good numerical stability.\\
We also applied the model to an idealized galaxy cluster merger to show that it is performant enough to be used in production runs.
The results of these simulations can be summarized as follows
\begin{enumerate}
    \item The subgrid modelling of CR proton injection is consistent with the lack of direct $\gamma$-ray observations in shocks due to the addition of shock obliquity dependent and the latest Mach number dependent injection models.
    \item Spectral treatment of CR electrons allows to calculate observables such as synchrotron emission directly from the time-evolved CR population.
    \item The injection models reproduce the complex radio relic morphology present in the \textsc{CIZA} relics.
    \item Ageing of the electron population leads to a steepening of the radio spectrum of the relic, which can be obtained directly from the simulation data.
    \item We can reproduce the shape of the synchrotron spectrum, albeit the absolute strength of the synchrotron emission is not reached. Cross-checking this against a shock tube test with realistic cluster shock parameters agrees well with previous results by other authors however, leading us to the opinion that the strength of the radio spectrum is most likely limited by the magnetic field strength over the whole relic.
\end{enumerate}
In upcoming work we will apply this model to high-resolution zoom-in simulations of galaxy clusters to study synchrotron and $\gamma$-ray emission of galaxy clusters.

\section*{Acknowledgements}
We thank the anonymous referee for their detailed and constructive feedback which improved the quality of this manuscript. 
We thank Beata Pasternak for preliminary work at the beginning of this project.
LMB would like to thank Denis Wittor, Franco Vazza, Micha\l $\:$ Hanasz, Paola Domínguez-Fernández, Mathias Hoeft, Vadim Semenov and Christoph Pfrommer for fruitful discussion.
LMB, KD and HL acknowledge support by the Deutsche Forschungsgemeinschaft (DFG, GermanResearch Foundation) under Germanys Excellence Strategy -- EXC-2094 -- 390783311. KD acknowledges support by the COMPLEX project from the European Research Council (ERC) under the European Union’s Horizon 2020 research and innovation program grant agreement ERC-2019-AdG 882679. The calculations were carried out at the Leibniz Supercomputer Center (LRZ) under the project pr86re. UPS is supported by the Simons Foundation through a Flatiron Research Fellowship at the Center for Computational Astrophysics of the Flatiron Institute. The Flatiron Institute is supported by the Simons Foundation.

\section*{Data Availability}

All data analysis scripts used for this work are available on Github at \url{https://github.com/LudwigBoess/2207.05087}. For the code tests these scripts contain a download step to obtain the data. For the cluster merger simulation the data will be provided by the corresponding author on reasonable request.

\section*{Software}
\textsc{P-Gadget3} \citep{Springel2005, Beck2016}, \textsc{Smac} \citep{Dolag2005}, \textsc{Julia} \citep{Bezanson2017}, \textsc{GadgetIO.jl} \citep[][]{GadgetIO}, \textsc{Matplotlib} \citep{Hunter2007}




\bibliographystyle{mnras}
\bibliography{references} 



\appendix

\section{Shocktube Parameters}
\begin{table*}
\centering
    \caption{We report the parameters for the different shock tubes tests used in this work. The subscripts $L$ and $R$ denote the quantities left and right of the initial contact discontinuity, respectively. From left to right we report model name, densities, temperatures, sonic Mach number in the case without CR acceleration and magnetic field vectors. We convert all units to physical units purely for consistency.}
\begin{tabular}{l|rrrrrrr}
\toprule
Model Name& $\rho_L$ [cm$^{-3}$] & $\rho_R$ [cm$^{-3}$] & $T_L$ [K] & $T_R$ [K] & $\mathcal{M}_{s,\mathrm{analytic}}$& $B_L$ [nG] & $B_R$ [nG]  \\
\midrule
\midrule
Sod$_{\mathcal{M}_s=3}$    & $356.18$ & $44.52$ & $4.75 \cdot 10^3$ & 745.6 & 3.00  & (0.0, 0.0, 0.0) & (0.0, 0.0, 0.0)  \\
Sod$_{\mathcal{M}_s=4}$    & $356.18$ & $44.52$ & $4.75 \cdot 10^3$ & 396.6 & 4.00  & (0.0, 0.0, 0.0) & (0.0, 0.0, 0.0) \\
Sod$_{\mathcal{M}_s=5}$    & $356.18$ & $44.52$ & $4.75 \cdot 10^3$ & 247.5 & 5.00  & (0.0, 0.0, 0.0) & (0.0, 0.0, 0.0) \\
Sod$_{\mathcal{M}_s=6}$    & $356.18$ & $44.52$ & $4.75 \cdot 10^3$ & 169.5 & 6.00  & (0.0, 0.0, 0.0) & (0.0, 0.0, 0.0) \\
Sod$_{\mathcal{M}_s=7}$    & $356.18$ & $44.52$ & $4.75 \cdot 10^3$ & 123.6 & 7.00  & (0.0, 0.0, 0.0) & (0.0, 0.0, 0.0) \\
Sod$_{\mathcal{M}_s=8}$    & $356.18$ & $44.52$ & $4.75 \cdot 10^3$ & 94.1 & 8.00  & (0.0, 0.0, 0.0) & (0.0, 0.0, 0.0) \\
Sod$_{\mathcal{M}_s=9}$    & $356.18$ & $44.52$ & $4.75 \cdot 10^3$ & 74.1 & 9.00  & (0.0, 0.0, 0.0) & (0.0, 0.0, 0.0) \\
Sod$_{\mathcal{M}_s=10}$   & $356.18$ & $44.52$ & $4.75 \cdot 10^3$ & 59.9 & 10.00 & (0.0, 0.0, 0.0) & (0.0, 0.0, 0.0) \\
Sod$_{\mathcal{M}_s=15}$   & $356.18$ & $44.52$ & $4.75 \cdot 10^3$ & 26.4 & 15.00 & (0.0, 0.0, 0.0) & (0.0, 0.0, 0.0) \\
Sod$_{\mathcal{M}_s=20}$   & $356.18$ & $44.52$ & $4.75 \cdot 10^3$ & 14.8 & 20.00 & (0.0, 0.0, 0.0) & (0.0, 0.0, 0.0) \\
Sod$_{\mathcal{M}_s=30}$   & $356.18$ & $44.52$ & $4.75 \cdot 10^3$ & 6.59 & 30.00 & (0.0, 0.0, 0.0) & (0.0, 0.0, 0.0) \\
Sod$_{\mathcal{M}_s=40}$   & $356.18$ & $44.52$ & $4.75 \cdot 10^3$ & 3.70 & 40.00 & (0.0, 0.0, 0.0) & (0.0, 0.0, 0.0) \\
Sod$_{\mathcal{M}_s=50}$   & $356.18$ & $44.52$ & $4.75 \cdot 10^3$ & 2.37 & 50.00 & (0.0, 0.0, 0.0) & (0.0, 0.0, 0.0) \\
Sod$_{\mathcal{M}_s=60}$   & $356.18$ & $44.52$ & $4.75 \cdot 10^3$ & 1.65 & 60.00 & (0.0, 0.0, 0.0) & (0.0, 0.0, 0.0) \\
Sod$_{\mathcal{M}_s=70}$   & $356.18$ & $44.52$ & $4.75 \cdot 10^3$ & 1.21 & 70.00 & (0.0, 0.0, 0.0) & (0.0, 0.0, 0.0) \\
Sod$_{\mathcal{M}_s=80}$   & $356.18$ & $44.52$ & $4.75 \cdot 10^3$ & $9.25 \cdot 10^{-1}$  & 80.00 & (0.0, 0.0, 0.0) & (0.0, 0.0, 0.0) \\
Sod$_{\mathcal{M}_s=90}$   & $356.18$ & $44.52$ & $4.75 \cdot 10^3$ & $7.31 \cdot 10^{-1}$  & 90.00 & (0.0, 0.0, 0.0) & (0.0, 0.0, 0.0) \\
Sod$_{\mathcal{M}_s=100}$  & $356.18$ & $44.52$ & $4.75 \cdot 10^3$ & $5.92 \cdot 10^{-1}$  & 100.00 & (0.0, 0.0, 0.0) & (0.0, 0.0, 0.0) \\
\midrule
Dubios$_{\theta = 0}$      & $356.18$ & $44.52$ & $4.53 \cdot 10^3$ & $57.0$ & $10.00$ & (0.1, 0.0, 0.0) & (0.1, 0.0, 0.0) \\
Dubios$_{\theta = 15}$     & $356.18$ & $44.52$ & $4.53 \cdot 10^3$ & $57.0$ & $10.00$ & (0.097, 0.026, 0.0) & (0.097, 0.026, 0.0) \\
Dubios$_{\theta = 30}$     & $356.18$ & $44.52$ & $4.53 \cdot 10^3$ & $57.0$ & $10.00$ & (0.087, 0.05, 0.0) & (0.087, 0.05, 0.0) \\
Dubios$_{\theta = 45}$     & $356.18$ & $44.52$ & $4.53 \cdot 10^3$ & $57.0$ & $10.00$ & (0.07, 0.07, 0.0)  & (0.07, 0.07, 0.0) \\
Dubios$_{\theta = 60}$     & $356.18$ & $44.52$ & $4.53 \cdot 10^3$ & $57.0$ & $10.00$ & (0.05, 0.087, 0.0) & (0.05, 0.087, 0.0) \\
Dubios$_{\theta = 75}$     & $356.18$ & $44.52$ & $4.53 \cdot 10^3$ & $57.0$ & $10.00$ & (0.026, 0.097, 0.0)  & (0.026, 0.097, 0.0) \\
Dubios$_{\theta = 90}$     & $356.18$ & $44.52$ & $4.53 \cdot 10^3$ & $57.0$ & $10.00$ & (0.0, 0.1, 0.0) & (0.0, 0.1, 0.0) \\
\midrule
Cluster                    & $1.81\cdot 10^{-4}$ & $5.14 \cdot 10^{-5}$ & $5.74 \cdot 10^8$ & $1.54 \cdot 10^{7}$ & 5.54  & ($3.5 \cdot 10^3$, 0.0, 0.0) & ($1.0  \cdot 10^3$, 0.0, 0.0) \\
\midrule
CIZA                       & $1.13\cdot 10^{-2}$ & $1.41 \cdot 10^{-3}$ & $5.58 \cdot 10^8$ & $3.46 \cdot 10^{7}$ & 4.60 & (0.0, 0.0, 0.0) & (0.0, 0.0, 0.0) \\
\bottomrule
\end{tabular}
\label{tab:shocktubes}
\end{table*}
We report the parameters for all shock tube tests used in this work in Table \ref{tab:shocktubes}.
The values are converted to physical units for clarity.
All tests are 3D setups of 140 stacked 1x1x1 boxes with 600 SPH particles in the r.h.s. of the contact discontinuity and 4800 SPH particles in the l.h.s.
This is the minimum resolution to avoid that the hydrodynamic smoothing length of the $C_6$ kernel with 295 nieghbors becomes larger than the box size and with that introduces double-counting of particles in the SPH loop.

\section{Performance and Scaling}
\label{app:performance_scaling}

\begin{figure}
	\centering
	\includegraphics[width=0.9\columnwidth]{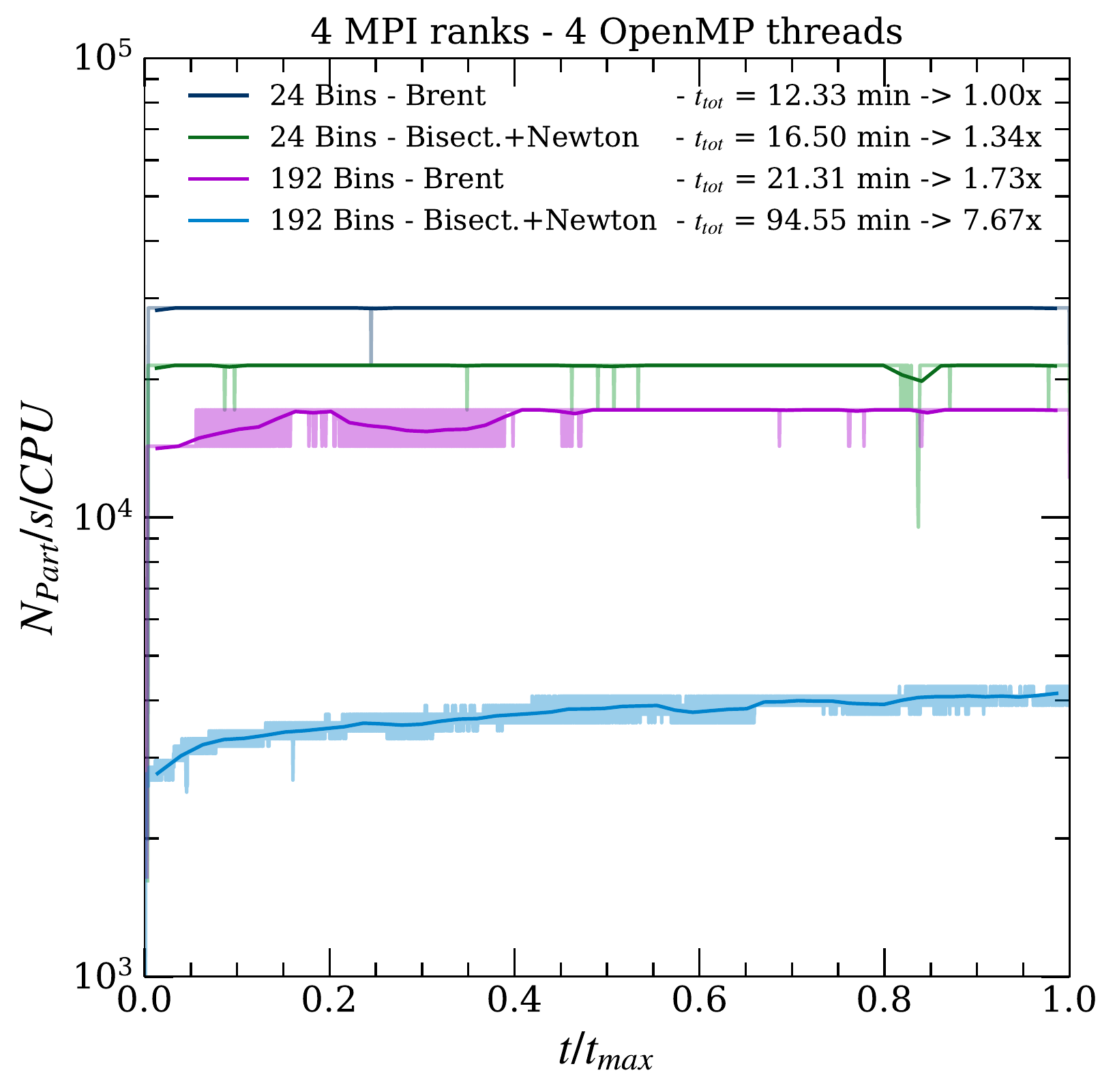}
	\caption{Performance comparison between different slope-solvers for a simple standalone cooling run. We compare the original method with an initial Bisection step to get a first guess for Newton's method and subsequent Newton's method step to the current implementation with Brent's method for 24 and 192 CR bins, respectively. Solid, dark lines indicate a fit to all timesteps shown with transparent lines.}
    \label{fig:slope_solver_performance}
\end{figure}

Fig.~\ref{fig:slope_solver_performance} shows the result of a simple scaling test for a small box of particles undergoing radiative cooling, with all other effects switched off. We compare our version with Brent's method to a legacy version using Bisection to get a first guess used for a subsequent Newton-Rhapson method. We find significantly better performance and more importantly better scaling using Brent's method.

\section{Decaying Sinewave} \label{appendix:sinewave}

\begin{figure*}
	\includegraphics[width=\fullwidth]{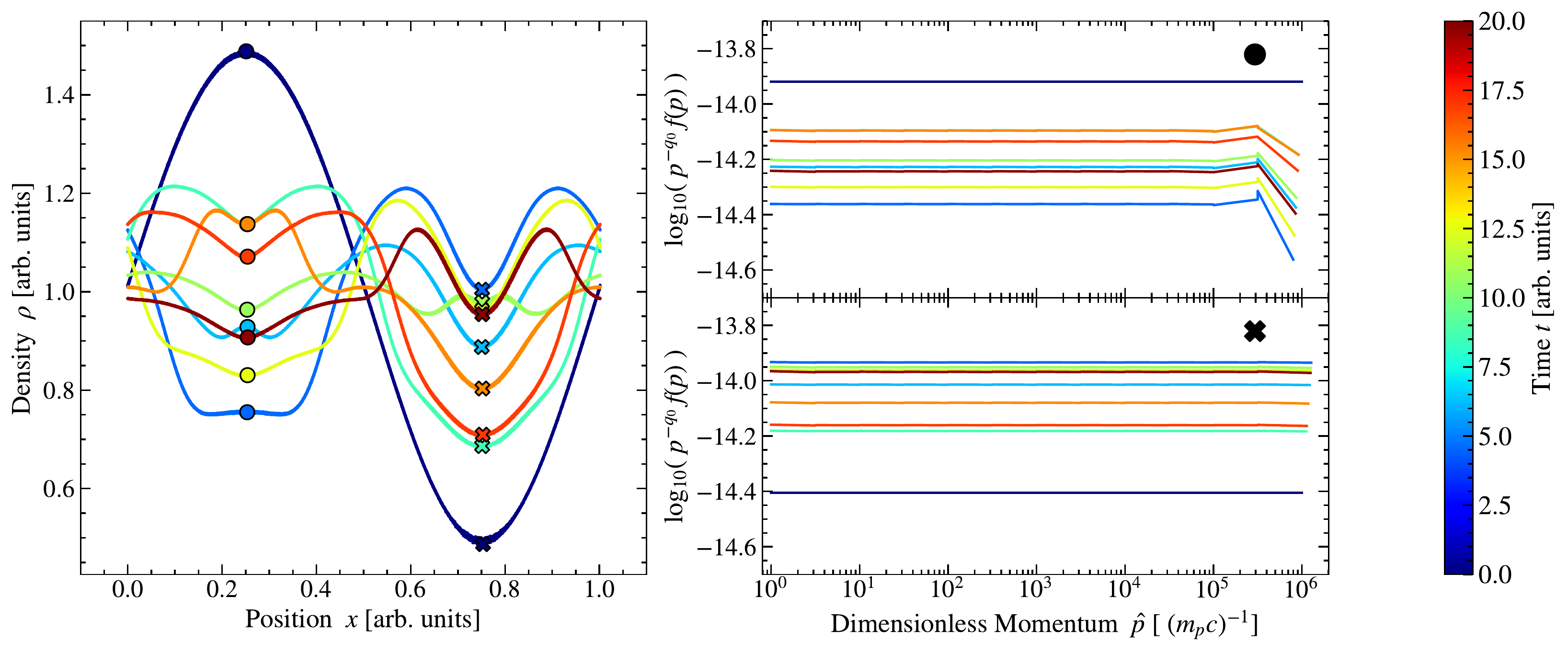}
    \caption{Test simulation of a sinus density profile in an isothermal periodic box. \textit{Left:} Time evolution of the density profile. \textit{Right:} Spectra of the particles with the maximum (upper panel) and minimum (lower panel) density at the beginning of the simulation. The corresponding particles are marked by the corresponding markers in the upper right corner of the spectrum plots. As desired the spectra stay flat over the course of the simulation with only the highest momentum bin showing a deviation under adiabatic expansion due to the spectral cut-off moving into the bin.}
    \label{fig:sine_spectra}
\end{figure*}

In order to test the stability of the adiabatic changes within \textsc{OpenGadget3} we set up a sinoidal density profile within an isothermal periodic box.
We set up a CR energy density as one third of the thermal energy density and let it evolve until the resulting pressure wave had completed approximately three modes.
The CR spectra were set up as a simple power-law with a constant slope $q_0 = -4.5$.
Fig.~\ref{fig:sine_spectra} shows the time evolution of the density in the left panel and the spectral evolution of two tracer particles in the right panels.
For the tracer particles we chose the particles with the highest density (large points) and lowest density (large X) at $t_0$ to observe the behaviour under primarily adiabatic expansion and compression, respectively.
We plot the particle distribution functions multiplied by $p^{q0}$ to visually emphasize discrepancies from the initial slope in a bin.
Since the distribution functions experience a self-similar shift due to adiabatic changes the slope should not change over the course of the simulation and should remain flat.
We see consistent flat spectra at the low momentum and only see deviation from this for the highest momentum bin in the particle that experiences adiabatic expansion.
This is expected as the spectral cutoff moves in the center of the bin.
Furthermore, it introduces only a small error, since the bulk of the energy is contained in the low-momentum end of the spectrum.
The stability of the distribution function is especially evident in the upper spectral panel for $t \approx 8.6$ (cyan line) and $t \approx 15.3$ (orange line).
As can be seen on the density plot on the left the density of the particle is virtually identical.
At the same times the spectral lines lie almost perfectly on top of each other, even though the particles have experienced one step of expansion and compression between the two data points.

\section{\textsc{CIZA} Shocktube}
\label{app:ciza_shocktube}
\begin{figure*}
    	\includegraphics[width=\fullwidth]{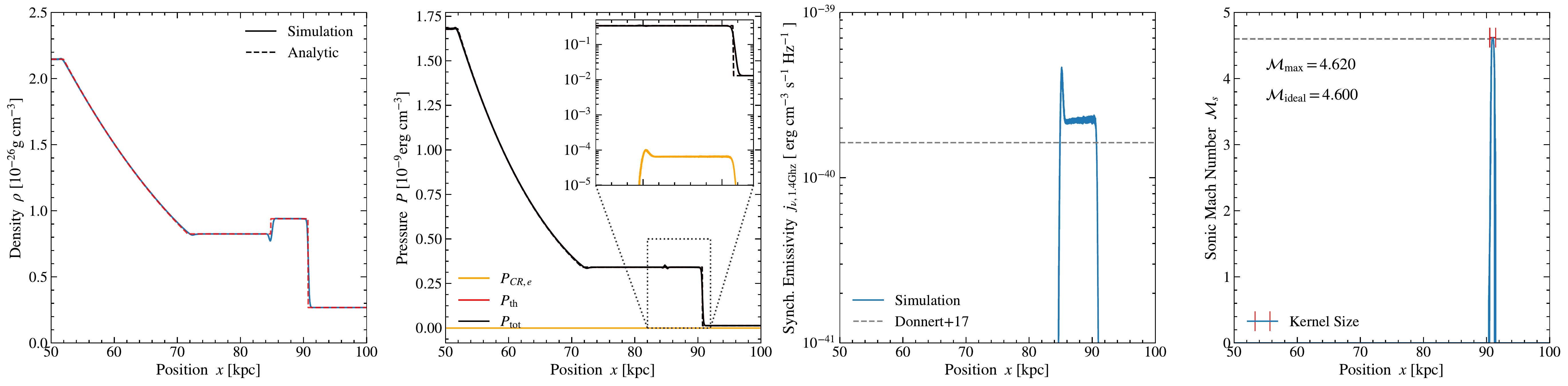}
        \caption{Idealized shock tube using the parameters for \textsc{CIZA} provided in \citet{Donnert2016}. We find excellent agreement to the analytic solution of the Riemann problem and shock capturing. For the synchrotron emissivity we employ a fixed magnetic field of $B = 5 \mu G$. This emissivity agrees well with the emissivity reported in \citet{Donnert2017}.}
        \label{fig:ciza_shocktube}
 \end{figure*}
We show the results of the \textsc{CIZA} shocktube discussed in Sec. \ref{sec:synch_power} in Fig. \ref{fig:ciza_shocktube}.
Initial parameters are given in Tab.~\ref{tab:shocktubes} and are obtained from \citet{Weeren2010, Ogrean2014, Akamatsu2015} as gathered by \citet{Donnert2017}.
We find overall excellent agreement with the analytic solution (dashed lines in density, pressure and Mach number), as well as the result for emissivity at 1.4GHz by \citet{Donnert2017}.

\section{Diffusion Times}
\label{app:diff}
\begin{figure}
	\centering
	\includegraphics[width=0.9\columnwidth]{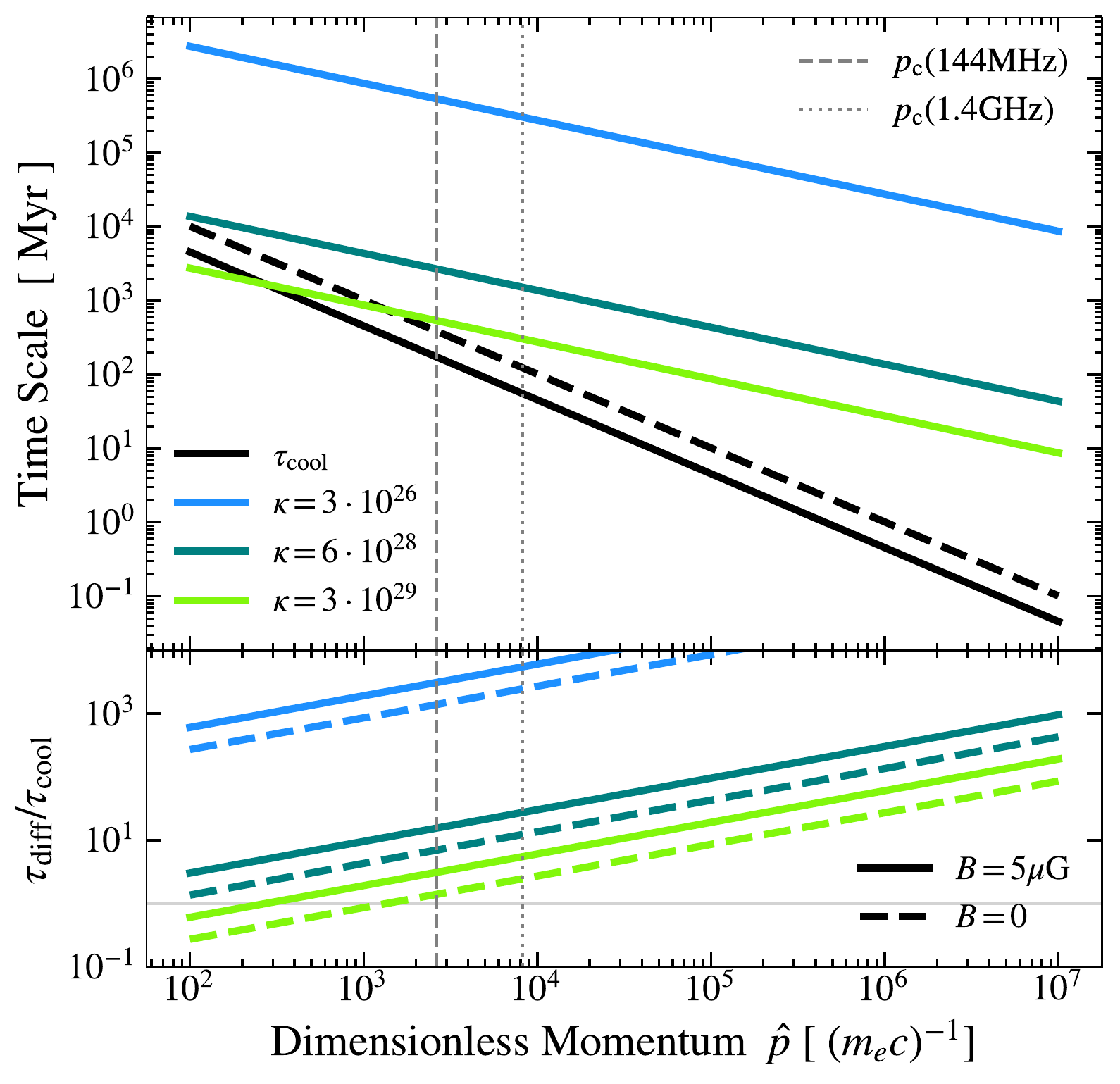}
	\caption{We show the comparison between diffusion (colors) and cooling times (black). The upper panel shows the absolute times, while the lower panel shows the time ratio. The vertical lines indicate the minimum dimensionless momentum of particles contributing to the synchrotron emission at the given observation frequencies.}
    \label{fig:diff_timescale}
\end{figure}
To better approximate the impact of a potential diffusion model on our results we employed a simple toy model to compare characteristic time-scales.
For this we assumed a momentum dependent diffusion coefficient as in \citet{Ogrodnik2020} of
\begin{equation}
    \kappa(\hat{p}) = \kappa_\mathrm{10k} \left( \frac{\hat{p}}{10^4} \right)^{\alpha_\kappa} 
\end{equation}
where $\kappa_\mathrm{10k}$ is the diffusion coefficient at $\hat{p} = 10^4$ and $\alpha_\kappa = 0.5$.
We approximate the diffusion timescale as $ t(p) \sim L^2 / \kappa(p)$ \cite[see e.g.][for analogous approach]{DAngelo2016}.
We assume that to have a significant impact on observations the electrons need to diffuse by at least the size of one radio beam and use $L = d_\mathrm{beam} \approx 16 \: \mathrm{kpc}$ for our adopted cosmology and the minimum resolution of $\theta_\mathrm{beam} = 5'$ given in \citet{Stroe2014}.
We apply this estimate to different values for $\kappa_\mathrm{10k}$, where we use $\kappa_\mathrm{10k} = 3 \cdot 10^{26} \frac{\mathrm{cm}^2}{\mathrm{s}}$ \citep[as in][]{Ogrodnik2020}, $\kappa_\mathrm{10k} = 6 \cdot 10^{28} \frac{\mathrm{cm}^2}{\mathrm{s}}$ \citep[][]{Trotta2011} and $\kappa_\mathrm{10k} = 3 \cdot 10^{29} \frac{\mathrm{cm}^2}{\mathrm{s}}$ \citep[][]{Chan2019}.
The result of which is shown as the colored lines in the upper panel of Fig. \ref{fig:diff_timescale}.
We compare this to  the cooling time of electrons due to synchrotron emission and IC scattering of CMB photons at $z=0$
\begin{equation}
    \tau_\mathrm{cool} =  \frac{3}{4} \frac{m_e^2 c^2}{\sigma_T (U_\mathrm{B} + U_\mathrm{IC}) p }
\end{equation}
for a magnetic field of $B = 5\mu$G and without magnetic field in the solid and dashed teal lines, respectively.
The lower panel shows the ratio of the respective timescales.
We mark the critical momenta for synchrotron emission \citep[see Eq. 22 in][]{Donnert2016} at 144 MHz and 1.4 GHz in the gray dashed and dotted lines.
Below these momenta the contribution of electrons to the total synchrotron emission drops off sharply.
We find that above these momenta the cooling times due to IC scattering alone dominates the evolution of the CRs.
Even in the extreme case where electrons can diffuse into regions of low magnetic field and the cooling time due to synchrotron losses increases, IC scattering still remains the dominant loss mechanism.
This picture may change in the context of fossil electrons as seeds for CR re-acceleration over cosmological timescales and has to be reconsidered in later work.


\bsp	
\label{lastpage}
\end{document}